\def\year{2021}\relax
\documentclass[letterpaper]{article} %

\usepackage{aaai21}  %
\usepackage{times}  %
\usepackage{helvet} %
\usepackage{courier}  %
\usepackage[hyphens]{url}  %
\usepackage{graphicx} %
\usepackage{natbib}  %
\usepackage{caption} %

\usepackage[show]{chato-notes}
\usepackage[framemethod=TikZ]{mdframed}
\usepackage{eucal}
\usepackage{booktabs}
\usepackage{multirow}
\usepackage{siunitx}
\usepackage{subcaption}
\usepackage{float}
\usepackage{xcolor}
\usepackage[linesnumbered,ruled,vlined]{algorithm2e}
\usepackage{graphicx}
\usepackage{pbox}
\usepackage{romannum}
\usepackage{soul}
\usepackage{pdfpages}
\usepackage{paralist}
\usepackage{amsfonts}
\usepackage{amsmath}

\newtheorem{observation}{Observation}

\newcommand{\spara}[1]{\smallskip\noindent\textbf{#1} }

\newcommand*{\mybox}[2]{\colorbox{#1!30}{\parbox{.89\linewidth}{#2}}}

\definecolor{mygray}{rgb}{0.75, 0.75, 0.75}

\begin{document}

\title{Exposure Inequality in People Recommender Systems: The Long-Term Effects}
\author{Francesco Fabbri,$^{1,2}$ Maria Luisa Croci,$^{3}$ Francesco Bonchi,$^{4,2}$ Carlos Castillo$^{1,5}$\\}

\affiliations{
$^{1}$Pompeu Fabra University, Barcelona, Spain --  email:  \{name.surname\}@upf.edu \\
$^{2}$Eurecat, Barcelona, Spain -- email: \{name.surname\}@eurecat.org \\
$^{3}$Sapienza University of Rome, Italy -- email: \{name.surname\}@gmail.com \\
$^{4}$ISI Foundation, Turin, Italy -- email:  \{name.surname\}@isi.it \\
$^{5}$ICREA, Barcelona, Spain.
}

\maketitle \sloppy

\begin{abstract}
People recommender systems may affect the exposure that users receive in social networking platforms, influencing attention dynamics and potentially strengthening pre-existing inequalities that disproportionately affect certain groups. 

In this paper we introduce a model to simulate the feedback loop created by multiple rounds of interactions between users and a link recommender in a social network. This allows us to study the long-term consequences of those particular recommendation algorithms.
Our model is equipped with several parameters to control:
\begin{inparaenum}[(i)]
\item the level of homophily in the network,
\item the relative size of the groups,
\item the choice among several state-of-the-art link recommenders, and
\item the choice among three different stochastic user behavior models, that decide which recommendations are accepted or rejected.
\end{inparaenum}

Our extensive experimentation with the proposed model shows that a minority group, if homophilic enough, can get a disproportionate advantage in exposure from all link recommenders. Instead, when it is heterophilic, it gets underexposed. Moreover, while the homophily level of the minority affects the speed of the growth of  the disparate exposure, the relative size of the minority affects the magnitude of the effect. Finally, link recommenders strengthen exposure inequalities at the individual level, exacerbating the\emph{``rich-get-richer''} effect: this happens for both the minority and the majority class and independently of their level of homophily.
\end{abstract}

\section{Introduction}
Contact recommender algorithms (e.g., \emph{``People You May Know''} in Facebook or \emph{``Who to Follow''} in Twitter) are recognized as key components in any on-line social networking platform: they help the users extending their network faster, thus driving engagement and loyalty \cite{ricci2011introduction,BarbieriBM14,xie2016learning,aiello}. As they affect the ``social capital'' --the number of followers one user has on these platforms-- these algorithms determine to a great extent the exposure and the amount of attention a user receives, potentially strengthening pre-existing inequalities and societal biases.
Effects such as the \emph{``algorithmic glass ceiling''} \cite{stoica} and more generally, inequalities in node rankings within and between groups of users (\citeauthor{karimi2018homophily}~\citeyear{karimi2018homophily}, \citeauthor{fabbri}~\citeyear{fabbri}), are phenomena that may be exacerbated by the recommenders, especially in presence of an homophilic behaviour.
Given that people recommenders are adopted not only in recreational social networks, but also in spheres such as
employment~\cite{HeapKWBC14,LiuORSXX16,LiuORSTX16,Ha-ThucVRSSG16,DomeniconiMPPP16,GuyRW09} and education~\cite{VassilevaMG16,ZhangMLSS16}, studying their potential biases is of great significance.

In our previous work we highlighted the harmful consequences of link recommenders after one round of recommendations (\citeauthor{fabbri}~\citeyear{fabbri}). Such a static picture can be limited as it does not study the consequential effects of the user behaviour which, by accepting or rejecting the recommendations, can determine the future structure of the social network and thus the exposure distribution. Specifically, multiple interactions between users and recommendation algorithms tend to nourish a \emph{feedback loop}: i.e., the output generated by the recommendation algorithm
is then fed as future input  for the next training of the recommender. In our setting, the recommended new links which are accepted, modify the structure of the network, thus constituting the input for the next cycle of link suggestions. In the context of items recommendation, recent simulation-based studies interested in the side-effects of collaborative filtering algorithms, show how a similar feedback loop \cite{feedback_loop-bias_amplification} impacts over user preferences, stimulating the \emph{popularity bias} \cite{beutel}. Those works underline the importance of providing models able to simulate the potential scenarios which may be otherwise difficult to investigate.

Following a similar approach, but focusing on people recommendations, in this paper we tackle the following research question: \emph{``which impact can link recommendation algorithms have over the network structure and user exposure along multiple rounds of recommendations?''}

\begin{figure}[t]
\includegraphics[width=.75\linewidth]{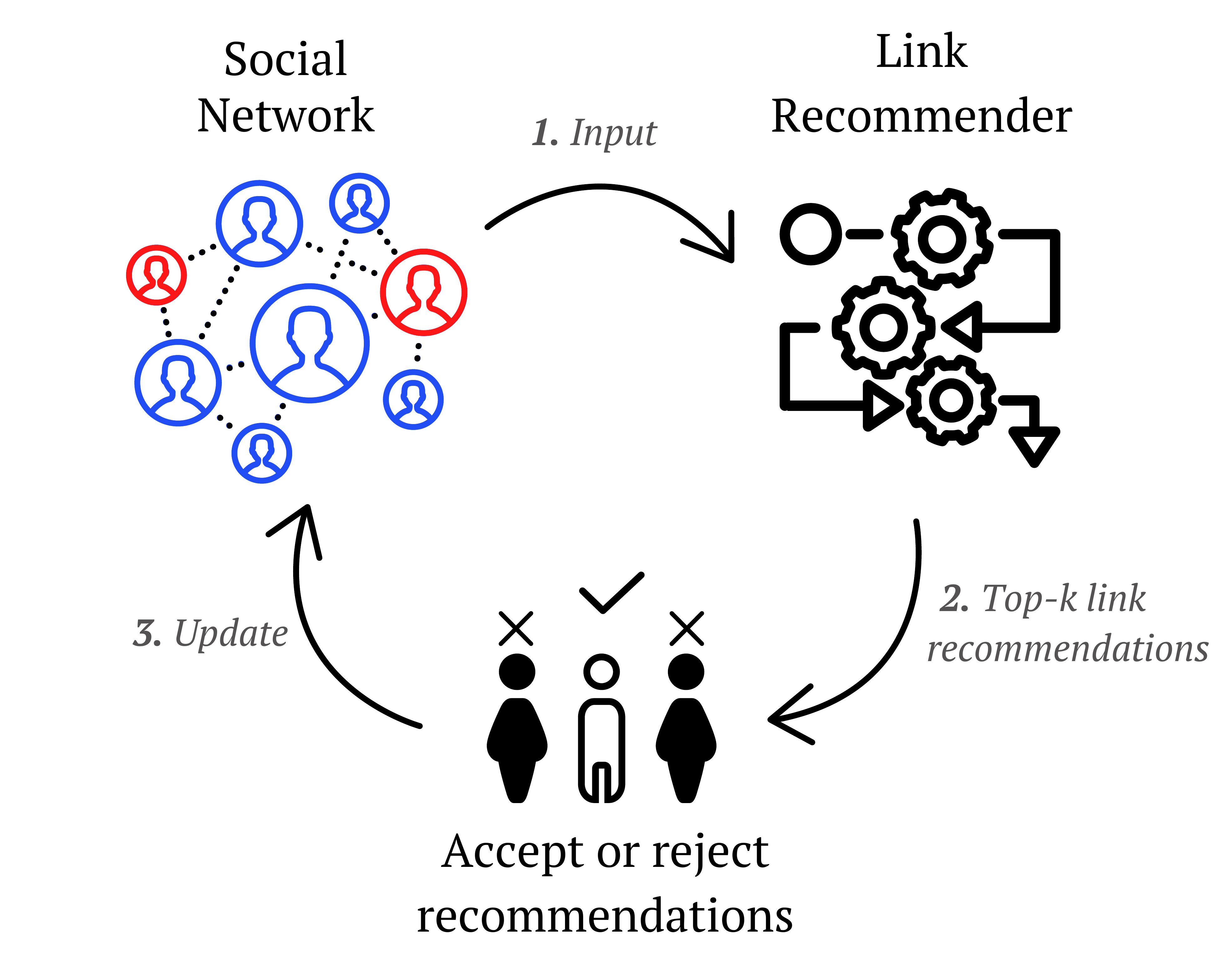}
\centering
\vspace{-2mm}
\caption{Bird's eye view of the simulation framework.}
\label{fig:framework}
\vspace{-4mm}
\end{figure}

Our contribution towards answering this question is a model able to simulate the long-term consequences of the injection of new recommended links into the network, reproducing the feedback loop triggered by the multiple interactions between users and the link recommender. Figure~\ref{fig:framework} provides a bird's eye view of our  proposed simulation model.

Starting with a social network where the nodes are partitioned in subgroups, e.g., by means of protected attributes such as gender or race (\S \ref{subsec:net}), a set of different link recommenders  are applied to the network to provide, at each iteration, $k$ link recommendations to each user (\S \ref{subsec:LR}). The user at this stage may then decide to accept or reject the recommendation. This decision is governed by three different stochastic
user behavior models (\S \ref{subsec:UBM}).
The rejected links are then discarded, while the new ones are included in the social graph.
The new augmented graph will then be the input to the next round of training of the recommender.
Despite our models does not consider the organic growth of the social graph, the simulations
 show that the injection of new links proposed by the recommendation algorithms can move the social graph far from the initial configuration.

\spara{Contributions and findings.} In this paper we propose a simulation model able to utilize several network configurations, user behaviors, and recommendation models in order to study the long-term effects of  people-recommender systems in social networks. We quantify the long-term disparate exposure generated by different initial network topologies, minority size and homophily level, and using different state-of-the-art link recommenders, and different
stochastic user behavior models. Our work confirms and extends the preliminary theoretical insights provided by \citeauthor{stoica}~(\citeyear{stoica}) and the empirical results of our previous work \cite{fabbri}, which was limited to one single round of recommendations.

Our findings are summarised as follows:

\begin{itemize}

\item Confirming the theoretical findings of \citeauthor{stoica}~(\citeyear{stoica}), our experiments show that, if the minority class is homophilic enough, it can get an advantage in exposure from all link recommenders. If the minority is heterophilic instead, it gets underexposed.

\item While the previous observation is robust to all the recommenders, the speed and magnitude of the disparate exposure along time differ across recommenders.

\item While the homophily of the minority affects the speed of the growth of disparate exposure, the size of the minority affects its magnitude.

\item The user behavior model (how recommendations are accepted or declined) does not impact significantly the evolution of exposure as much as the initial network configuration and the algorithm do.

\item Some recommenders can strengthen exposure inequalities at the individual level: after a few iterations, most of the links are recommended towards a small subset of ``super-star'' nodes. This happens for both the minority and the majority class and independently of their level of homophily. Hence, in the long-term, the ``rich-get-richer'' effect is exacerbated .
\end{itemize}

\section{Related Work}

In this section we discuss the literature most related to our work. We divide the presentation into two topics: work dealing with inequalities in social networks, and simulation-based studies in recommender systems.

\spara{Inequalities in social networks.} In our previous work \cite{fabbri} we observed, in a ``static'' single round of recommendations, that homophily is a driving force in shifting visibility distribution. In particular, we introduced the concept of disparate visibility in a bi-populated network, showing how effects such bias in rankings and rich-get-richer can get amplified by homophilic networks. The main limitation of our previous study is that it looks at one single round of recommendations, missing the long-term effects.

\citeauthor{stoica}~(\citeyear{stoica}) study the existence of a \textit{glass-ceiling} (an invisible barrier that prevents women from rising in higher rankings) effect amplified by the combination of organic growth and a random walk recommenders. Their preliminary analytical findings inspired us to study, through simulations, different recommenders and distinct network topologies. The main limitation of \cite{stoica} is that  its theoretical findings require many strong assumptions to hold, such as,  the existence of a power inequality between groups in-degree distributions and the use of a specific, random-walk based, recommender.

\citeauthor{leeetal}~(\citeyear{leeetal}) show that the perceptions about the size of minority groups in social networks can be biased, often exhibiting systematic over- or underestimation. Moreover, these biases can be explained by the level of homophily and by the size of the minority class. Our work, is inspired by their insights, extending the analysis of the inequalities while the network is injected with new links driven by the recommendation output. We confirm their observations, showing how the recommender algorithms can introduce even more inequality along the time.

\citeauthor{tsioutsiouliklis2020fairness}~(\citeyear{tsioutsiouliklis2020fairness}) propose methods for fairness-aware link analysis, introducing techniques able to mitigate unfairness generated by Pagerank. Later, in \S \ref{sec:results},  we will show  that another popular random-walk based recommender (i.e., SALSA) can increase the unfairness in visibility in the long run, thus confirming the need to devise methods able to mitigate these effects.

\spara{Simulation-based studies in recommender systems.}
\citeauthor{cinus22}~(\citeyear{cinus22}) combine link recommendation and opinion-dynamics
models in a simulation-based framework, to assess the effect of people recommenders on the evolution
of opinions in a social network. They show that, if the initial network exhibits high level of homophily, people recommenders can help creating echo chambers and polarization.

In the context of collaborative-filtering-based methods, \citeauthor{feedback_loop-bias_amplification}~(\citeyear{feedback_loop-bias_amplification}) show that popularity bias can be stimulated by feedback loop, where popular items tend to obtain more and more interactions if generated through recommendations. 
\citeauthor{jiang2019degenerate}~(\citeyear{jiang2019degenerate}) propose a theoretical framework to model the effects of ``\textit{filter bubble}'', i.e., the tendency of the recommendation algorithm to drive the preferences of the user towards a limited amount of items. 
Recently, \citeauthor{beutel}~(\citeyear{beutel}) propose a simulation model for measuring the impact of recommender systems over time, analyzing the changes in the user experience with an application designed for food recommender system.

Our work is motivated by the importance of studying algorithmic bias in recommendations and rankings
in the long term, i.e.,  beyond the single round of algorithmic intervention.
In this regard, \citeauthor{Towards_Long_term_Fairness}~(\citeyear{Towards_Long_term_Fairness}) have recently introduced the problem of \emph{long-term fairness}, designing also solutions able to account for algorithmic unfairness in the long-term in movies recommendations.
\citeauthor{gini}~(\citeyear{gini}) propose a simulation model able to include multiple recommender systems combined with different users choice models, proving that the rich-get-richer effect tends to increase over time, stimulated by the algorithm. In our study we analyze the evolution of rich-get-richer effect in social networks, fueled by the edges created thanks to the recommendation algorithms.

\citeauthor{nguyen2014exploring}~(\citeyear{nguyen2014exploring}) show how in the case of MovieLens data, recommendations generated through a collaborative filtering approach have not strengthen the filter bubble effect. In our paper we go in the opposite direction, showing how homophily may generate biased recommendations, towards a smaller set of recommendations, reducing the diversity of those. \citeauthor{sun2019debiasing}~(\citeyear{sun2019debiasing}) propose a mitigation strategy to reduce popularity bias in recommendations through different methods based on active-learning. The methods proposed are aimed at reducing popularity bias, which in our setting can be related to rich-get-richer effect. Although this paper may be used to propose mitigation strategies to reduce inequality in exposure, the main weakness of their work regards the lack of analysis of different input data distribution. This kind of analysis may help to understand which kind of distribution of user-item interactions may benefit more from their method.

\section{Model}
We consider a social graph whose nodes are partitioned by demographics (e.g. gender, age or other characteristics). More formally, let $G =(V,E,\ell)$ be the social graph, where $V$ is the set of nodes, $E \subseteq V \times V$ is the set of directed edges, such that an edge $(u,v) \in E$ indicates the fact that $u$ \emph{follows} $v$.
Finally $\ell: V \rightarrow \{V_m, V_M\}$ is a labeling function assigning each node to either the minority ($V_m$) or the majority ($V_M$) class (with $|V_m| < |V_M|$). We denote by $s_m = |V_m|/|V|$ the fraction of nodes belonging to the class less represented in the network, i.e., the \emph{minority}, and by $s_M$ the fraction of nodes belonging to the majority

\spara{Homophily.} To capture the bias in the distribution of the edges towards each group, we introduce a measure of homophily, expressed as \emph{the tendency of people in a a group to connect to individuals in the same group}. We model the homophily as the portion of edges distributed within the same group discounted by the fraction observed in a random configuration \cite{fabbri}. More formally:
\begin{equation}\label{eq:h}
h_i = \frac{|E_{ii}|}{|E_{i.}|} - s_i
\end{equation}

\noindent
where $E_{ii} = \{(u,v) \in E | u \in V_i \wedge v \in V_i\}$ and  $E_{i.} = \{(u,v) \in E | u \in V_i\}$.
This measure ranges in the interval $(-s_i,1-s_i]$. A group is called \emph{homophilic} if the tendency to connect to nodes of the same group is stronger than expected ($h_i>0$), \emph{heterophilic} when this tendency is weaker than expected ($h_i<0$), and \emph{neutral} if the number of edges within the group is comparable to the relative size of the group ($h_i = 0)$.

\begin{figure*}[h]
\centering

\includegraphics[width=1\linewidth]{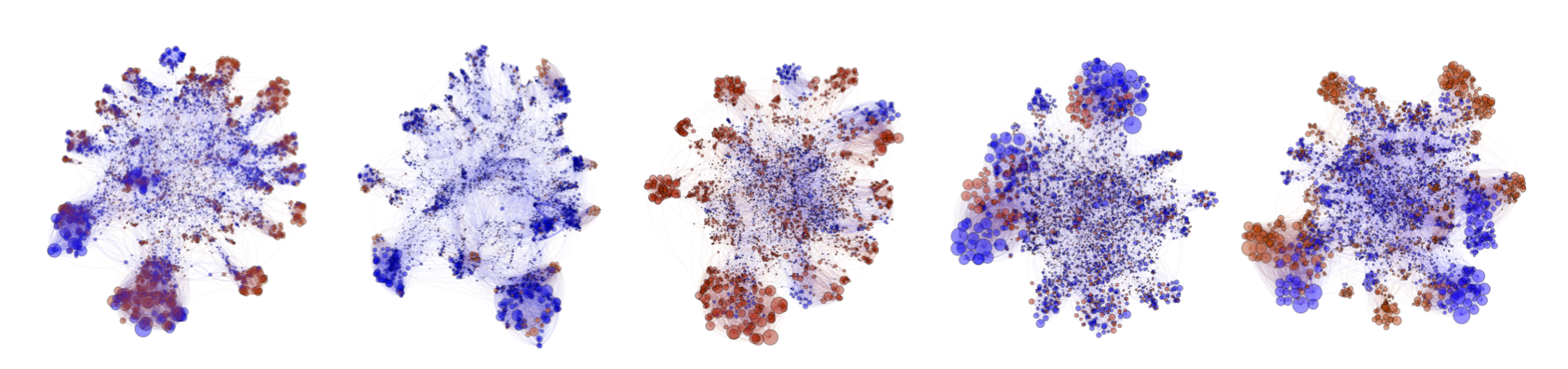}

\resizebox{\linewidth}{!}{\begin{tabular}{ccccc}

\begin{subfigure}[t]{.20\linewidth}\centering (a) G0 \\(original)\end{subfigure}
&
\begin{subfigure}[t]{.20\linewidth}\centering (b) G1 \\(homophilic minority)\end{subfigure}
&
\begin{subfigure}[t]{.20\linewidth}\centering (c) G2 \\(one homophilic group)\end{subfigure}
&
\begin{subfigure}[t]{.20\linewidth}\centering (c) G3 \\(heterophilic minority)\end{subfigure}
&
\begin{subfigure}[t]{.20\linewidth}\centering (d) G4 \\(two homophilic groups)\end{subfigure}

\end{tabular}
}

\caption{Representation of a sample of each generated network, where the minority is indicated in red, while the majority in blue. Each sample considers 5,000 nodes and those are the ones with highest degree in each group. Specifically for each group, a total of $s_{i}\times$ 5,000 ($i \in \{m, M  \}$) nodes are sampled.}
\label{fig:dataset_gephi}
\end{figure*}

\spara{Step-by-step.}
In our simulations we reproduce the multiple interactions between the users and the recommendation algorithm, where at each round, a set of new links is recommended to a portion of users randomly sampled from the graph. This sampling represents the fact that only a set of users are online at a certain time, helping reproducing a more realistic scenario.
Then the users accepts or rejects the recommendations according to a given stochastic user behavior model. This process is iterated for a given number of iterations.The graph grows accordingly to the new accepted recommendations: neither organic growth, nor edge removal are considered.
Table \ref{tab:steps} summarizes the simulation process step-by step.

\begin{table}[t!h!]
   \caption{Simulation steps.}\label{tab:steps}
\vspace{-3mm}

\begin{mdframed}[innerbottommargin=3pt,innertopmargin=3pt,innerleftmargin=6pt,innerrightmargin=6pt,backgroundcolor=gray!10,roundcorner=10pt]
\smallskip
\begin{enumerate}
\item \textbf{Input.} We start with an initial network configuration with specified levels of homophily and size of the minority class (how this initial configuration is generated by modifying a real-world social graph is presented in \S \ref{subsec:net}). We also set parameters such as the number of recommendations $k$ that a user receives in a round,  the number of iterations $T$, the fraction $\alpha$ of users to sample, the link-recommendation algorithm $A$ (presented in \S \ref{subsec:LR}), and the stochastic user behavior model $B$ (discussed in \S \ref{subsec:UBM}).

\item \textbf{Recommendation round.} A link recommender model is trained over the current social graph by the algorithm $A$. A portion $\alpha$ of users is sampled from the network. Those sampled users receive their top-$k$ recommendations each. The recommendations are links never recommended before and are generated from the set of missing edges at distance two (e.g. ``friends of friends'').

\item \textbf{Graph update.} Each user decide to accept or reject each of the $k$ recommended links, according to the model $B$. The social graph is thus updated by adding the newly accepted links. Each link rejected at this stage is discarded and never recommended again.

\item \textbf{Repeat.} Steps $2$ and $3$ are repeated $T$ times.
\end{enumerate}
\smallskip
\end{mdframed}
\vspace{-3mm}
\end{table}

For all the results that we report in \S \ref{sec:results}
we use $T = 20$, $\alpha = 20\%$ and $k=3$.

We next present in more details the various key components of our model.

\subsection{Initial network configuration}\label{subsec:net}
In order to control the level of homophily and the size of the minority class, while keeping a realistic network structure, we propose a novel data generation process which, starting with a real-world bi-populated network, performs just the minimum amount of node class-swappings and link rewirings to match the requested levels of homophily and size of the minority class. In this regard, our networks are \emph{semi-synthetic}. The process is explained in details next.

Our starting real-world network comes from Tuenti, a social network popular few years ago in Spain, which was known as ``the Spanish Facebook''. The dataset includes demographic information about users as gender and age (\citeauthor{LaniadoVKK16}~\citeyear{LaniadoVKK16}). The network has 8,983,560 nodes (users) and 17,830,103 edges, where a generic edge $(u,v)$ indicates a user posting on another users' \emph{wall}.
Along the paper, we use a sample of the original network used in our previous work \cite{fabbri}. This sample, which is also the one from which we derive the other configurations, is given by partitioning the users by age (16 as cut-off). We call it \textbf{G0}: it contains 500,000 nodes and 2,813,744 edges, with a relevant minority ($s_{m} = 0.30$) and both groups (majority and minority) appearing to have some level of homophily. Starting from this network, we generate 4 different semi-synthetic networks, ranging different values of $h$ and $s$.
More in details, let $V_m$ and $V_M$ be respectively the set of minority's and majority's node in the input network and $N_m = |V_m|, N_M = |V_M|$. Let also  $h_m, h_M$ indicate the actual homophily of each class, and finally $s_m$ and $s_M$ are the relative size of both classes. The generation process takes as input the desired level of homophily for both classes, denoted $h^*_m, h^*_M$ and the desired proportion for the minority class  $s_{m}^{*}$, and works as follows:

\begin{enumerate}

\item \textbf{Change minority-majority size.}  Let $N^*_m = (N_m + N_M) s_{m}^{*}$.
 If $N^*_m < N_m$, then $N_m - N_m^*$ nodes are sampled at random from the minority class $V_m$ and their label is flipped to the majority. Otherwise, we sample of $N_m^* - N_m$ nodes are extracted from the majority $V_M$ and their label flipped to the minority.

\item \textbf{Change homophily.}  For each group i $\in \{m, M\}$ we first compute the difference between the initial and the final homophily $| h^*_i - h_i| = B_{i}$. Then depending on the sign of the difference $ h^*_i - h_i$ we define which edges need to be rewired. Rewiring an edge $(u,v)$ means substituting $(u,v)$ with an edge $(u,w)$ such that $\ell(v) \neq \ell(w)$.
    If $ h^*_i - h_i > 0$ a sample of edges beloning to $E_{ij} = \{(u,v) \in E | u \in V_i \wedge v \in V_j\}$ is selected and rewired towards nodes in $V_{i}$. In this way we increase the set of nodes in $E_{ii}$, reaching the requested level of homophily. Viceversa, if  $ h^*_i - h_i > 0$ the operation is the opposite: old edges belonging to the subset $E_{ii}$ are rewired towards nodes in $V_{j}$. In both cases, the final amount of edges rewired is $E_{i.} \times B_{i}$.

\end{enumerate}

\begin{table}[t]
\caption{Table summarizing information about the generated graphs. For each one we have: i) name, ii) scenario characterizing the network } \label{tab:networks}
\resizebox{\linewidth}{!}{
\begin{tabular}{cccc}
\toprule
Graph & Scenario  & $s_m$ & $h_m$  \\
\midrule
G0 & \emph{original} & 0.3  & 0.42   \\
G1  &  \emph{different sizes + homophilic minority} & 0.1  & 0.4     \\
G2   & \emph{same sizes + homophilic minority} & 0.45 & 0.5   \\
G3    & \emph{different sizes + heterophilic minority} &  0.3  & -0.25    \\
G4  & \emph{different sizes + homophilic groups}& 0.3  & 0.6  \\
 \bottomrule
\end{tabular}
}
\end{table}

Since some recent literature has shown that small subpopulations within a social network can impact the whole graph \cite{stoica, fabbri, karimi2018homophily}, we generate networks with biased distributions for the minorities. Only for one case, to have a comprehensive analysis, we modify the homophily level  in both minority and majority groups.

More in details, these are the configurations we focus on:

\begin{itemize}
\item \textbf{G1.} To analyze the effect of a \textbf{small homophilic minority} we generate a graph with $s_{m} = 0.1$ and $h_{m} = 0.4$, with a neutral majority. %

\item \textbf{G2.} To emphasize the role of homophily we also generate a graph with \textbf{comparable sizes} between the two groups $(s_{m} = 0.45)$ with the minority strongly homophilic $(h_{m} = 0.5)$ and a majority still neutral. %

\item \textbf{G3.} This configuration is the unique with a \textbf{small heterophilic minority} $(h_{m} = -0.25)$ (and neutral majority).

\item \textbf{G4.} The the final configuration has \textbf{both subpopulations homophilic}. In particular, we keep a small minority $(s_{m} = 0.3)$ and both groups with high level of homophily ($h_{m} = 0.6$ and $h_{M} = 0.2$). %
\end{itemize}

$G1$ and $G2$ are a useful comparison against $G0$, since they present comparable level of homophily but different sizes of the minority, while $G3$ is useful to explore the  heterophilic case and $G4$ resembles a scenario quite common in contexts where phenomena such as polarization and filter bubbles drive the network formation \cite{garimella2018quantifying}. Table \ref{tab:networks} summarizes the five networks used in our analysis, while Figure~\ref{fig:dataset_gephi} depicts a sample of each network.

\subsection{Link recommenders}\label{subsec:LR}
Link recommendation algorithms are selected accordingly to state-of-the-art performance and popularity in the literature (\citeauthor{li2017survey}~\citeyear{li2017survey}, \citeauthor{fabbri}~\citeyear{fabbri},  \citeauthor{stoica}~\citeyear{stoica}). In accordance with our data model, all the methods recommend directed links (i.e., who to follow).

\begin{description}
	\item{\textbf{ADA: Network Topology Based.}} Among the different heuristics which aim to define similarity between nodes looking at the graph topology, we select the directed version of the \emph{Adamic-Adar coefficient} (for short ``ADA'' in the rest of the paper), method that penalizes connections with high degree nodes \cite{dada}.

	\item{\textbf{SLS: Random Walks Based.}} As representative of random-walks based approaches, we use SALSA (Stochastic Approach for Link-Structure Analysis; ``SLS'' in the rest of the paper), which is at the basis of the \emph{who-to-follow} recommender at Twitter \cite{goel1}. Recommendation of a generic link is defined as the probability of the source node to jump to the target one, rather than to any other node in the graph.

	\item{\textbf{ALS: Collaborative Filtering Based.}} Connections among nodes can be considered as implicit feedback in a collaborative filtering approach. An Alternating Least Squares algorithm (``ALS'' in the rest of the paper) is selected to perform recommendations \cite{hu2008collaborative}. New links are suggested based on latent features extracted from the adjacency matrix.

	\item{\textbf{RND: Random baseline.}} As baseline, we consider a random recommender (``RND'' in the rest of the paper), which picks recommendations uniformly at random from the candidate nodes at distance 2.
\end{description}

\subsection{User behavior models}\label{subsec:UBM}
In order to simulate the user feedback on the received recommendation, we consider three stochastic user behavior models. The first two are adapted from a recent work simulating user-item interactions \cite{beutel}, while the third one defines acceptance probability biased by the position in the ranked list of recommendations. Through these stochastic choice models the users add in expectation one edge per recommended list.

\begin{itemize}
\item \textbf{B-LZY} - \emph{Lazy}. The user accepts directly the first recommendation:
\begin{equation*}
 P(\emph{u selects v at position i} ) =
  \begin{cases}
 1  \quad \text{if  i = 1} \\
 0  \quad \text{otherwise}
 \end{cases}
\end{equation*}

\item \textbf{B-RND} - \emph{Random}. The user picks in the top-k, one recommendation uniformly at random:
$$ P(\emph{u selects v at position i}) = \frac{1}{k} $$

\item \textbf{B-PSB} - \emph{Position Biased}. This policy refers to the idea of having user choices biased by the position bias of rankings, where the user may accept or reject the recommendation with probability based on its position \cite{position_bias}. Hence, higher ranked suggestions are more likely to be chosen:

$$ P(\emph{u selects v at position i}) =  \frac{1/\log{(i + 1) }}{ \sum_{j = 1}^{k} 1/\log{(j + 1)}}. $$

\item \textbf{B-MIX} - \emph{Mixed}. In order to evaluate how heterogenous user behaviors may affect the exposure distribution, we also include B-MIX, which is a combination of the previous policies. Specifically, at each iteration, each user first picks, uniformly at random, one of the three strategies above, then follows it.

\end{itemize}

\noindent B-PSB model resembles the classical position bias, observed as key factor for predicting clickthrough rates in search engines \cite{position_bias, joachims2002optimizing}. The other two user behavior resembles two extreme situations: i) B-RND is the one less dependent by the order of the recommendation list; while B-LZY represents the case in which the user relies completely on the order imposed by the recommendation algorithm.

\section{Results}\label{sec:results}

In this section we present the results of our experiments, focusing on the key measure that we call \emph{exposure} of the minority,  which is simply defined as \emph{the portion of total number of recommendations which suggest a node of the minority}, and denoted $\mathcal{E}_{m}$. Note that the total number of recommendations is constant and corresponds to  $k|V^{\alpha}|$.

\subsection{Exposure in the long-run}
\begin{figure}[t!]

\centering
\includegraphics[width=0.9\linewidth]{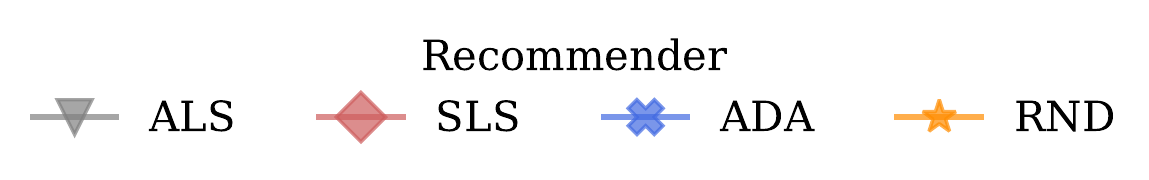}\\

\begin{tabular}{cc}
\centering

             \hspace{-3mm}\includegraphics[width=0.5\linewidth]{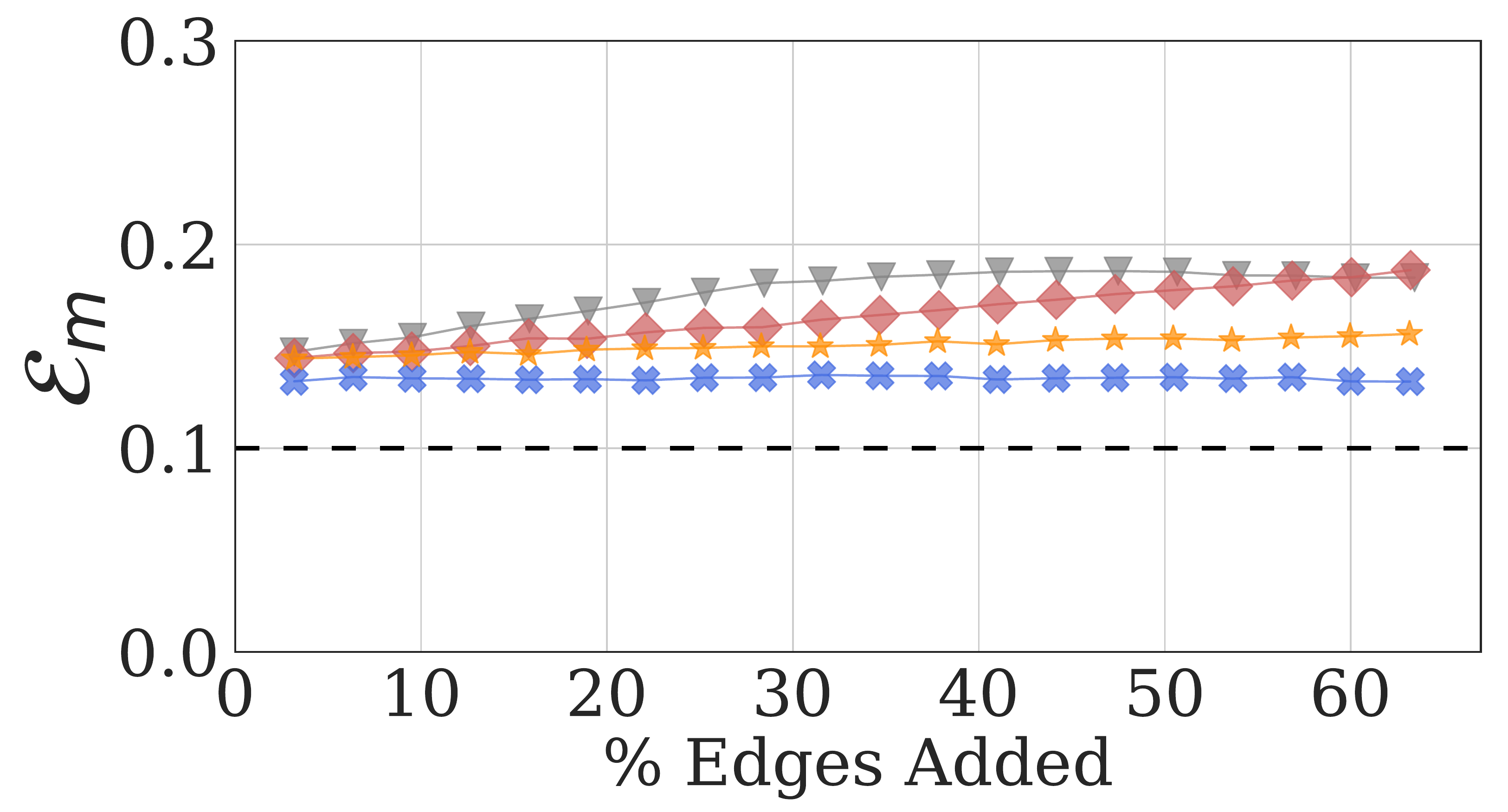}&
             \hspace{-3mm}\includegraphics[width=0.5\linewidth]{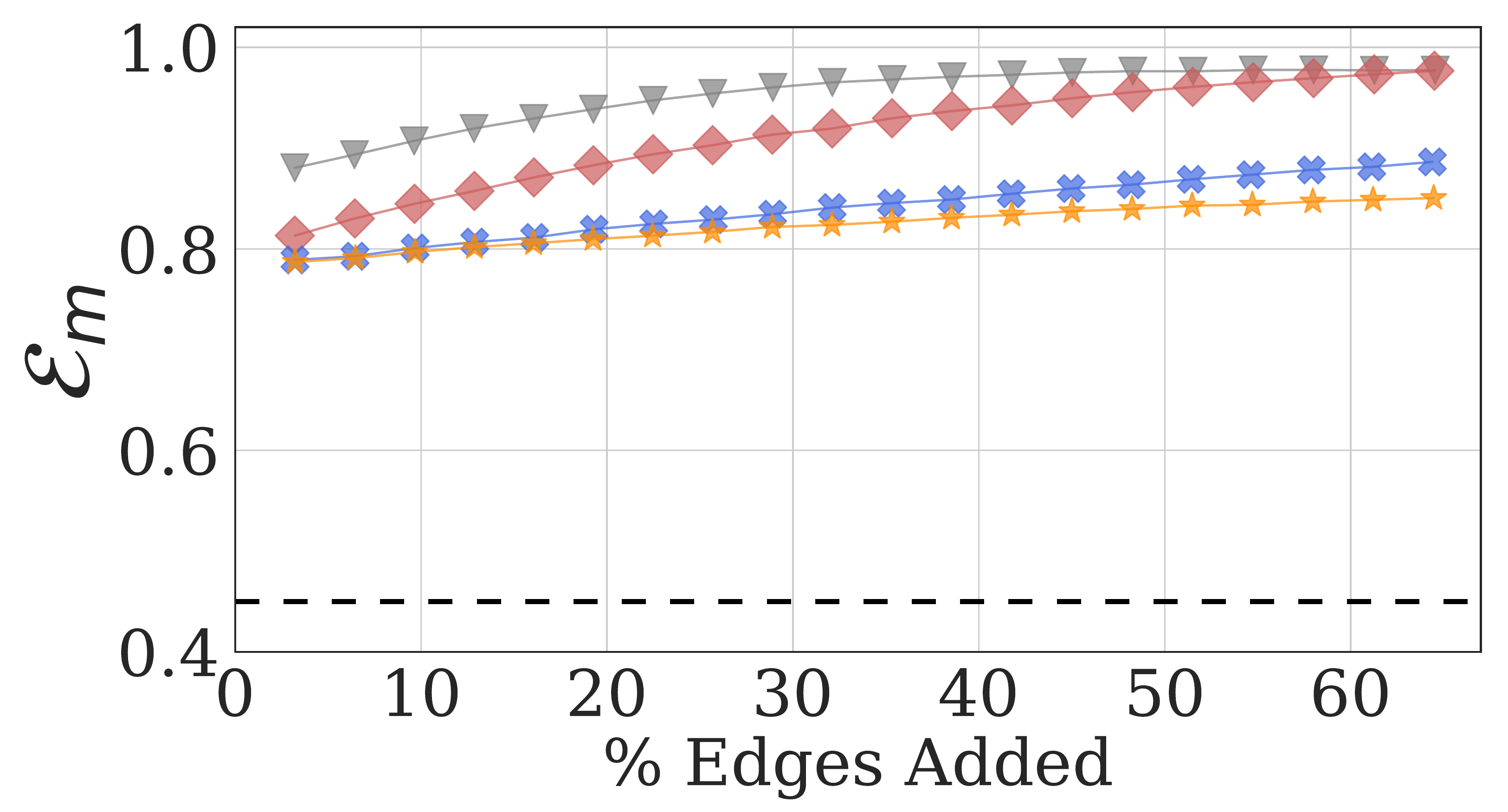}\\

	    (a) G1 & (b) G2 \\
	    \hspace{-3mm}\includegraphics[width=0.5\linewidth]{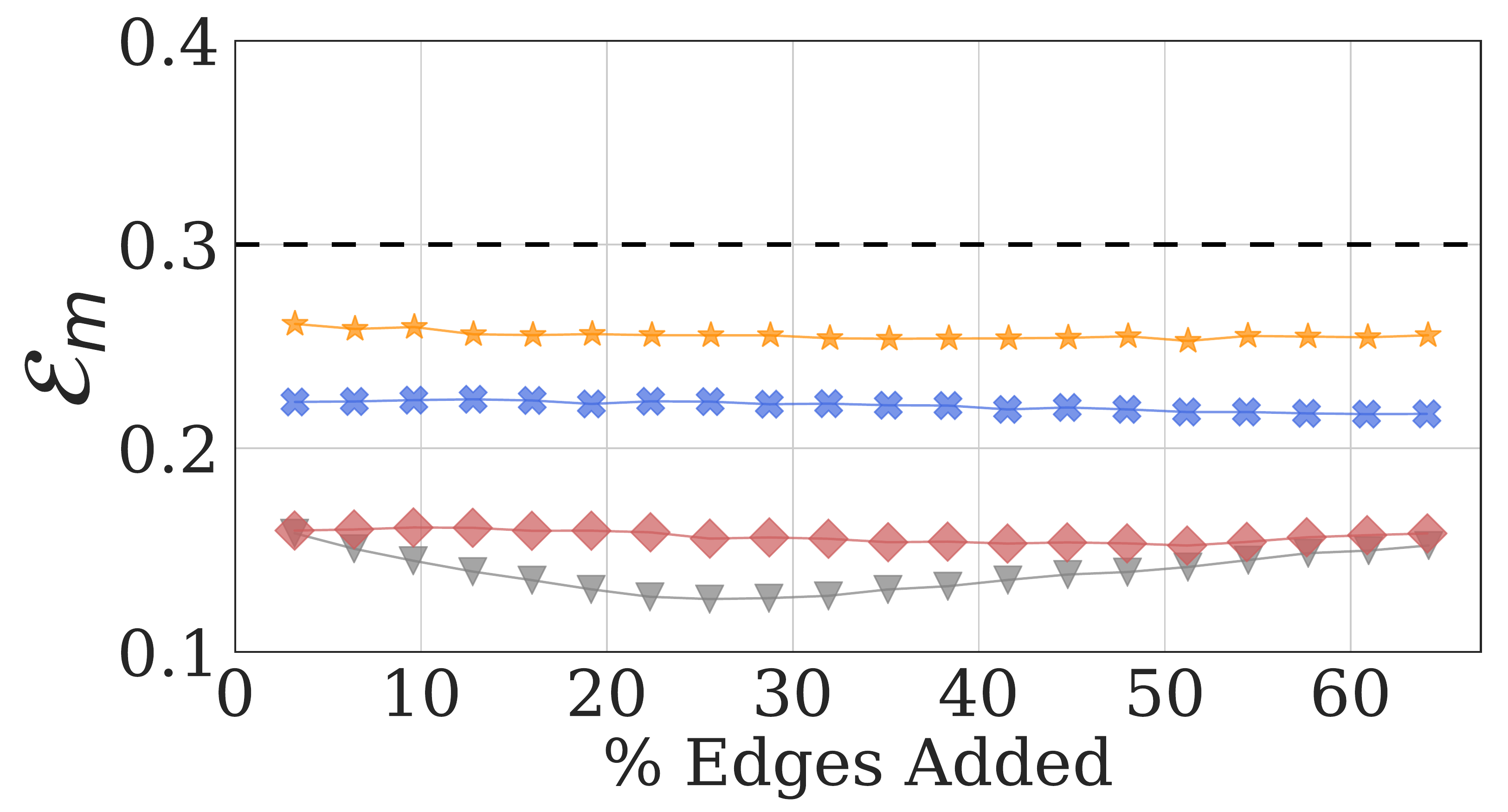}&
	    \hspace{-3mm}\includegraphics[width=0.5\linewidth]{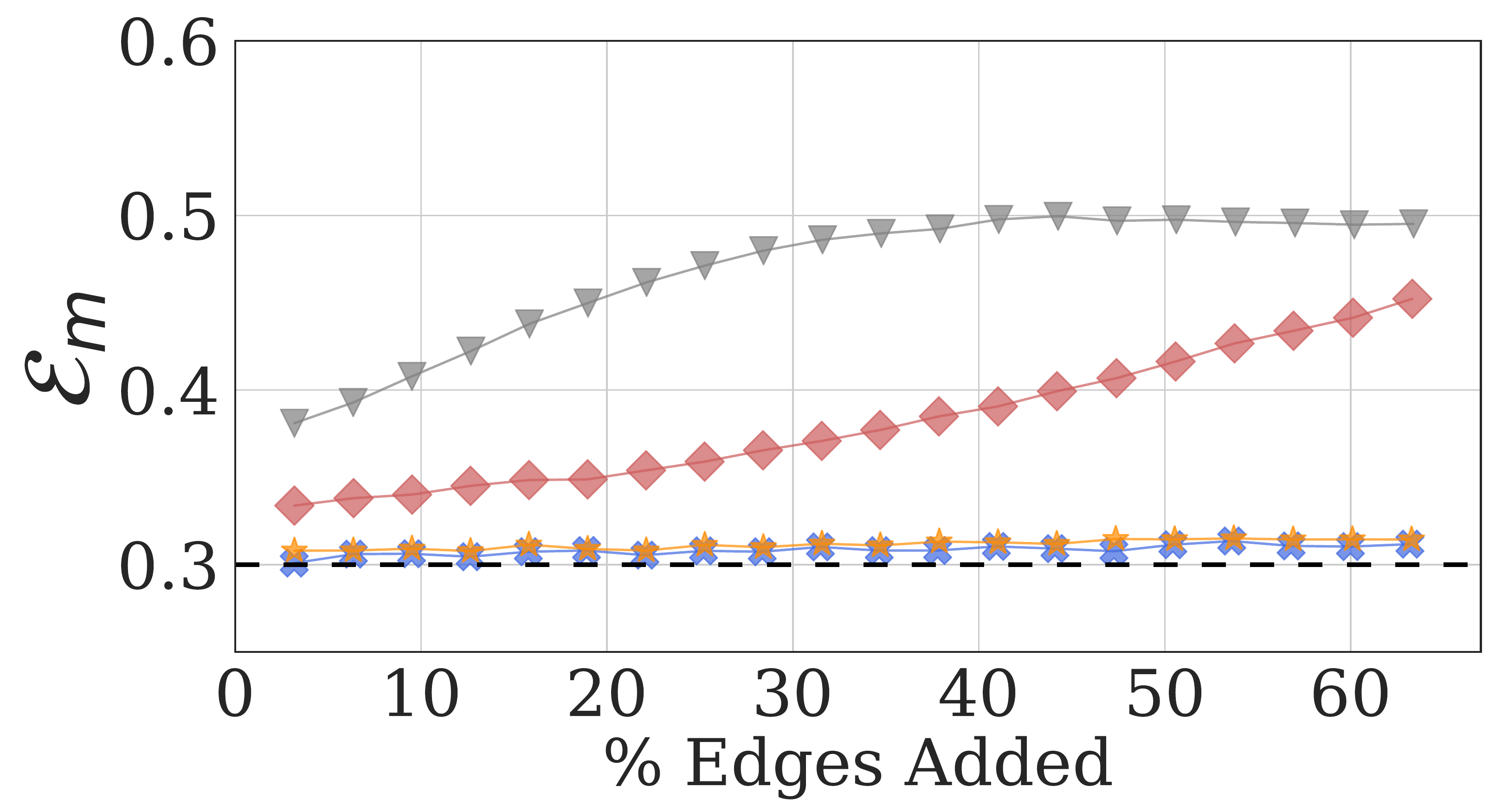} \\
	
	    (c) G3 & (d) G4
\end{tabular}

\caption{Exposure of the minority ($\mathcal{E}_{m}$) along time, for different recommenders and one fixed user behavior (B-PSB).}
\label{fig:exposure}
\end{figure}

Figure \ref{fig:exposure} shows the trend of the exposure of the minority for each of the four networks (G0 is omitted for space limitation as it presents results almost indistinguishable from G4). For each experiment, we track the exposure and the percentage of new edges added at each iteration with respect the original network. The dashed line represents in each plot the relative size of the minority and the user choice model is fixed to B-PSB. In cases when the minority is homophilic (G1, G2 and G4), generating recommendations through ALS and SALSA leads to a positive trend of growth for the exposure of the minority. For the other two recommenders (ADA and RND) the effects described above are still present but less visible.

For the case in which the minority is heterophilic (G3) the exposure decreases weakly, slightly benefitting the majority. This is the only case when the exposure distributed to the minority is less than its relative size. It is also evident how the collaborative filtering approach (ALS) and the random walk based model (SLS) contribute more to reduce the exposure allocated to the minority with respect the other two models (ADA and RND).
For all the networks characterized by an homophilic minority (G1, G2 and G4), the growth in the case of the collaborative filtering approach (ALS) is faster in the first steps and then stabilizes to a steady-state to the rest of iterations. While, for the random walk solution (SLS), the trend starts at similar values of exposure, but then grows constantly.

\smallskip

\mybox{mygray}{ \begin{observation} The disparate exposure grows after each iteration in favour of the minority, when it is homophilic. On the other hand, an heterophilic behavior of the minority does not impact abruptly its exposure. When both groups are homophilic, the recommender still increases the exposure of the minority. The severity of all those effects is stronger when using ALS and SLS and weaker for ADA and RND.
 \end{observation}
 }

\smallskip

\begin{figure}[h!]
    \centering
\begin{tabular}{cc}
            \hspace{-8mm}\includegraphics[width=0.62\linewidth]{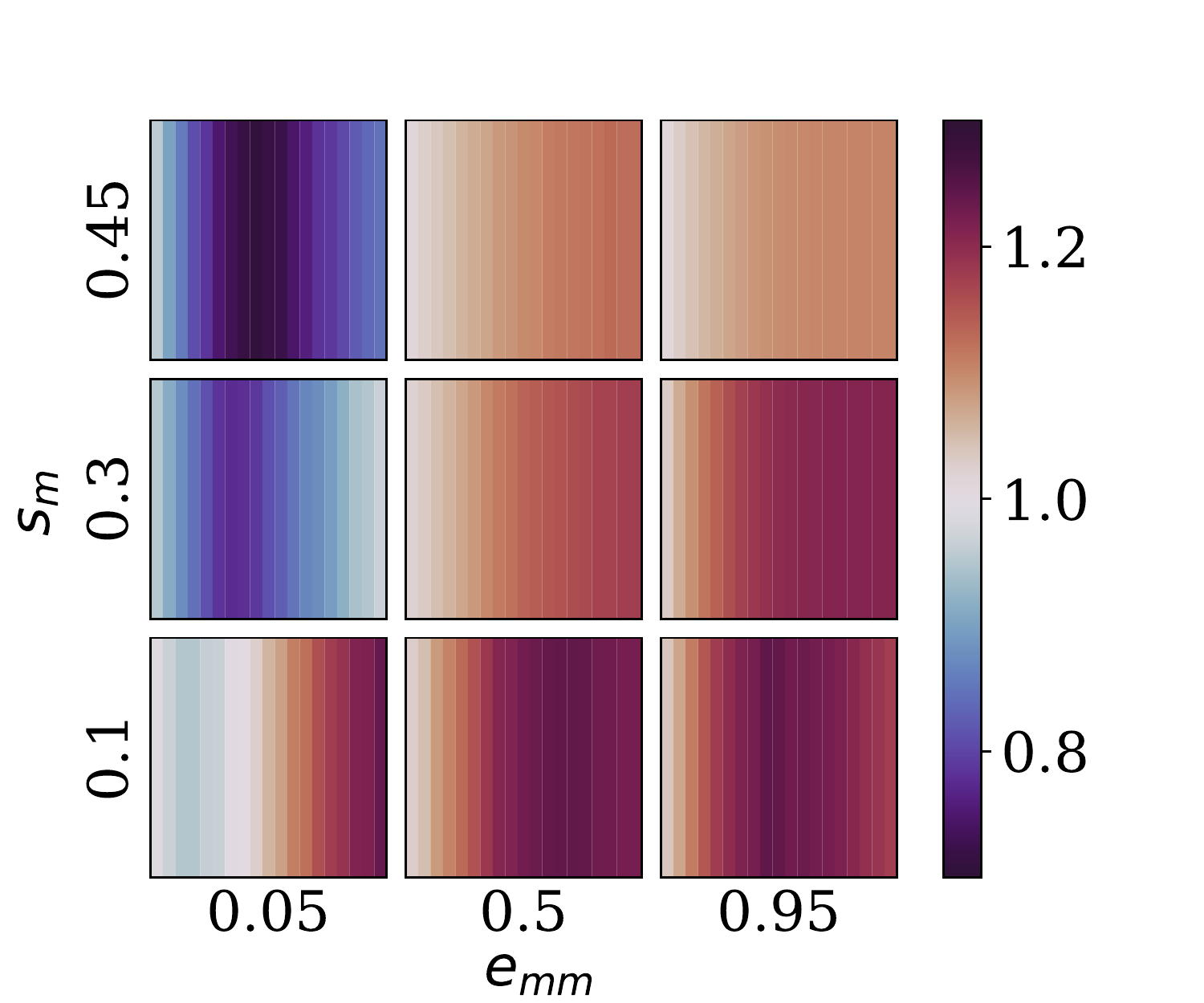} &
            \hspace{-8mm}\includegraphics[width=0.62\linewidth]{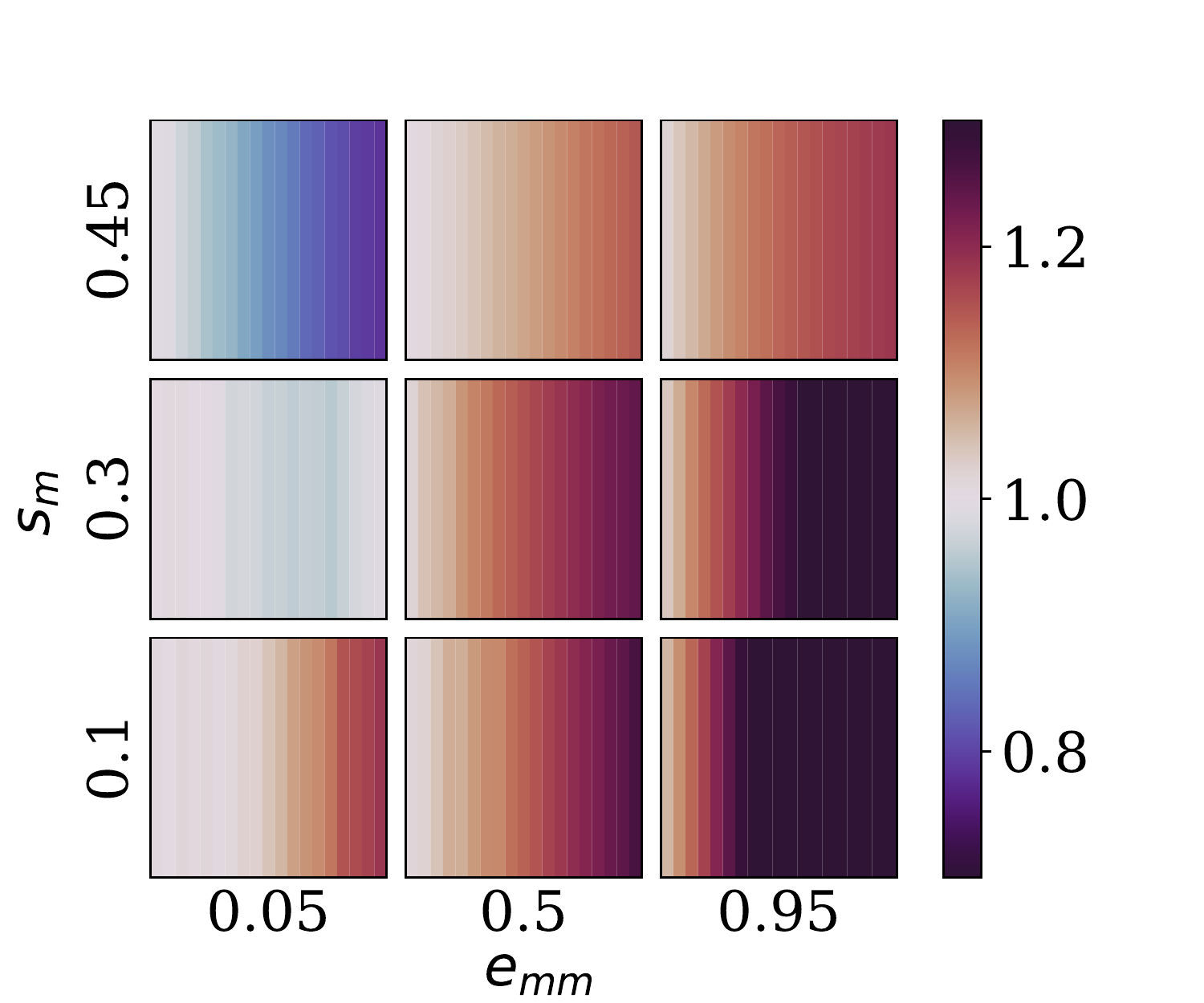}\\
            \hspace{-12mm}ALS & \hspace{-12mm}SLS\\
\end{tabular}
\vspace{-2mm}
\caption{Heatmaps describing the evolution of $\mathcal{E}_{t}/ \mathcal{E}_{1}$ in $T=20$ iterations, computed over 9 configurations which are small variants of G1, G2 and G3:
$e_{mm} \in \{ 0.05, 0.5, 0.95\}$  ($x$-axis) and $s_{m} \in \{ 0.1, 0.3, 0.45\}$ ($y$-axis), all having neutral majority $h_M = 0$. ALS recommender (left-hand side), SLS recommender (right-hand side).}
\label{fig:heatmaps}
\end{figure}

\begin{figure*}[t]
\centering
\begin{tabular}{c}
\hspace{-5mm}\rotatebox[origin=c]{90}{$\mathcal{E}_{t}/ \mathcal{E}_{1}$\hspace{-80mm}}\includegraphics[width=1.\textwidth]{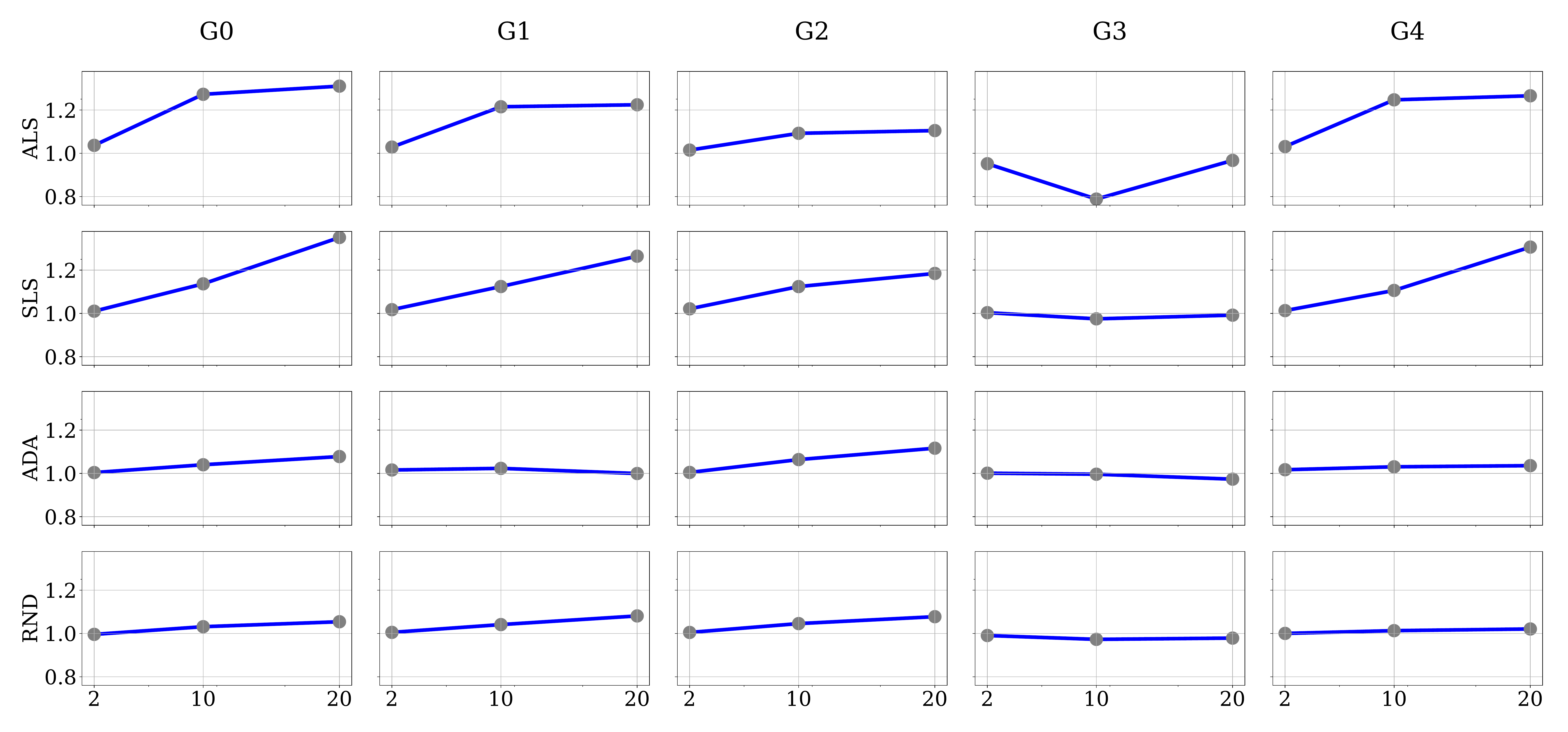}
\end{tabular}
Iteration
\caption{Evolution of exposure relative to the one observed at first iteration $\mathcal{E}_{t}/ \mathcal{E}_{1}$, after 1, 10 and 20 iterations.}
\label{fig:growth-rate}
\end{figure*}

To further investigate the differences in growth of exposure, we track $\mathcal{E}_{t}/ \mathcal{E}_{1}$ for $t \in \{2, ..., T\}$, which is relative quantity of exposure measured with the respect to the first iteration. In order to analyze the transition phases previously mentioned, we focus respectively on the iterations $t \in \{2, 10, 20\}$.
In Figure \ref{fig:growth-rate} we plot those values on the $y$-axis and the iterations on the $x$-axis. As suggested by the previous plots, with the ALS recommender, when the minority is homophilic its exposure tends to increase faster in the first iterations to then stabilize.
On the other hand, SLS presents a continuous increase, without slowing down the process after 20 iterations. The stronger growth comes from cases where the differences in sizes between minority and majority is relevant $(s_{m} = 0.3)$ and the minorities are homophilic (G0, G1 and G4).
This means that, even when also the majority is homophilic, the effect is still present, showing again that having both groups homophilic does not imply a benefit for the majority.
As already seen in Figure \ref{fig:exposure} and as expected, ADA and RND do not produce much exposure disparity, even a slight advantage for the minority class can be observed for the cases in which the minority is homophilic.

\smallskip

\mybox{mygray}{ \begin{observation}
Different recommenders exhibit different influence on exposure along time. ALS increases exposure inequality in the first iterations, then stabilizing in a steady state. SLS instead keeps increasing disparate exposure constantly.
\end{observation}
}

\smallskip

We further extend this analysis, exploring a wider range of initial configurations, with the aim of disentangling the effects of size and homophily along time. For this purpose, as the size of the minority $s_m$ is part of the definition of the homophily $h_m$ (Eq. 1), we use directly the fraction of edges which, starting from the minority, remain in the minority: i.e.,
$$
e_{mm} = \frac{|E_{mm}|}{|E_{m.}|}
$$

In particular, we produce 9 configurations which are small variants of G1, G2 and G3:
$e_{mm} \in \{ 0.05, 0.5, 0.95\}$ and $s_{m} \in \{ 0.1, 0.3, 0.45\}$ (all having neutral majority $h_M = 0$).

Each box of the heatmaps in Figure \ref{fig:heatmaps} represents, for one configuration, the evolution of $\mathcal{E}_{t}/ \mathcal{E}_{1}$ along $T=20$ iterations.

Analyizing the two heatmaps, comparing the boxes by columns and posing the attention on a single row, we observe that both effects already observed in the previous experiments, i.e. the steady-state generated by ALS and the constant growth caused by SLS, change in terms of severity but not in timing. This means that the variation of $\mathcal{E}_{t}/ \mathcal{E}_{1}$ can be less or more severe, depending on the distribution of $e_{mm}$, but the pace to which process evolves is the same. Analyzing the heatmaps by rows, and posing the attention on a single column, is evident how the size of the minority (represented here by $s_{mm}$) can influence the pace of the effects but not the range of values (color intensity) of  $\mathcal{E}_{t}/ \mathcal{E}_{1}$.

\smallskip
\mybox{mygray}{ \begin{observation} The homophily of minority can impact the speed at which the growth of exposure disparity occurs. On the other hand, the severity of this effect is mostly determined by the size of the minority.
\end{observation}
}

\smallskip

\subsection{Effect of user behavior models}
\begin{figure}[t]
    \vspace{2mm}
    \centering
    \includegraphics[width=0.8\linewidth]{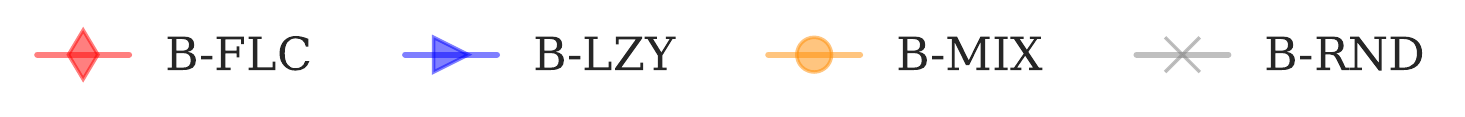}
    \begin{tabular}{cc}        \hspace{-3mm}\includegraphics[width=0.5\linewidth]{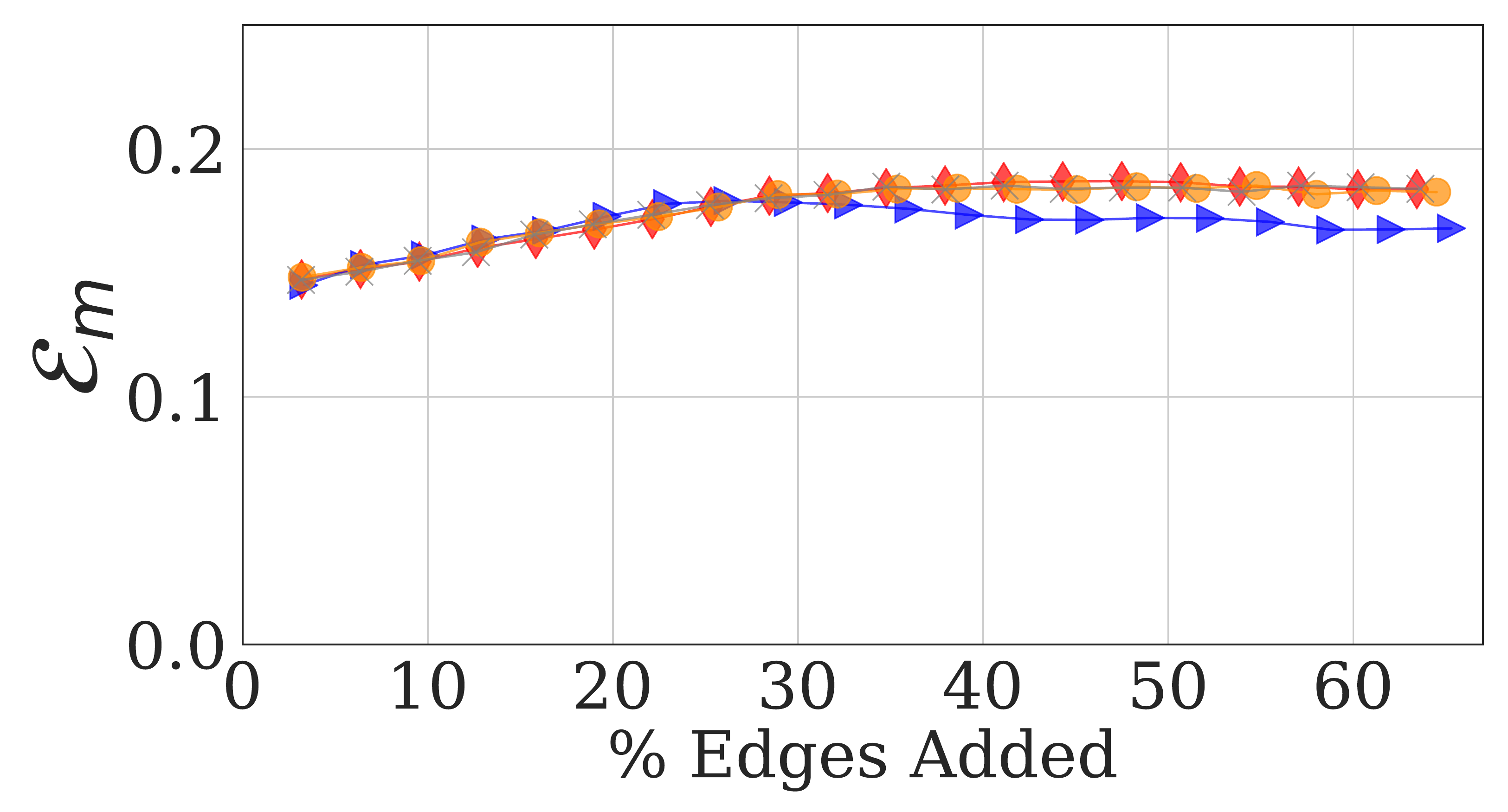} &
        \hspace{-3mm}\includegraphics[width=0.5\linewidth]{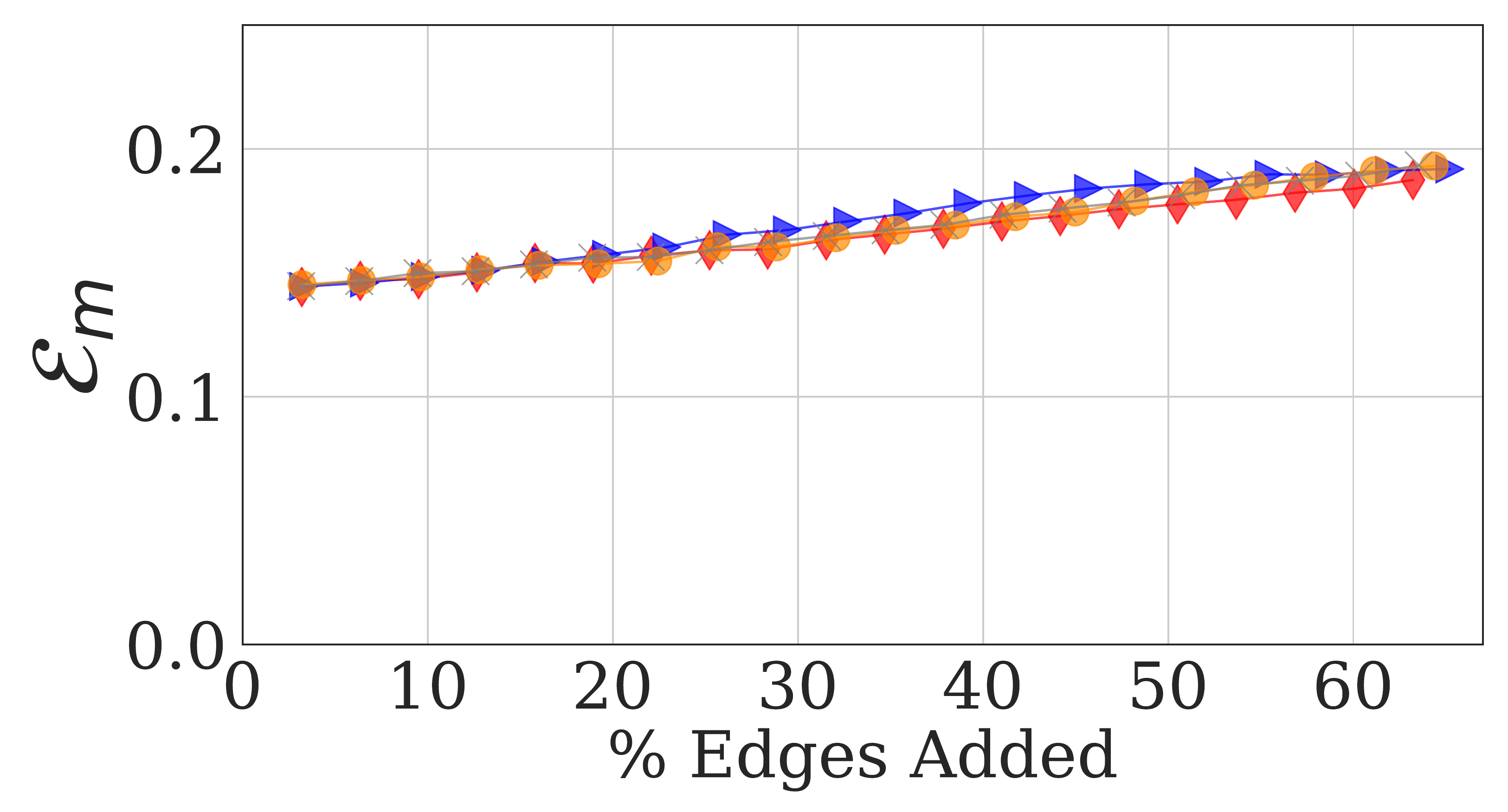}\\
        (i) ALS & (ii) SLS\\
        \hspace{-3mm}\includegraphics[width=0.5\linewidth]{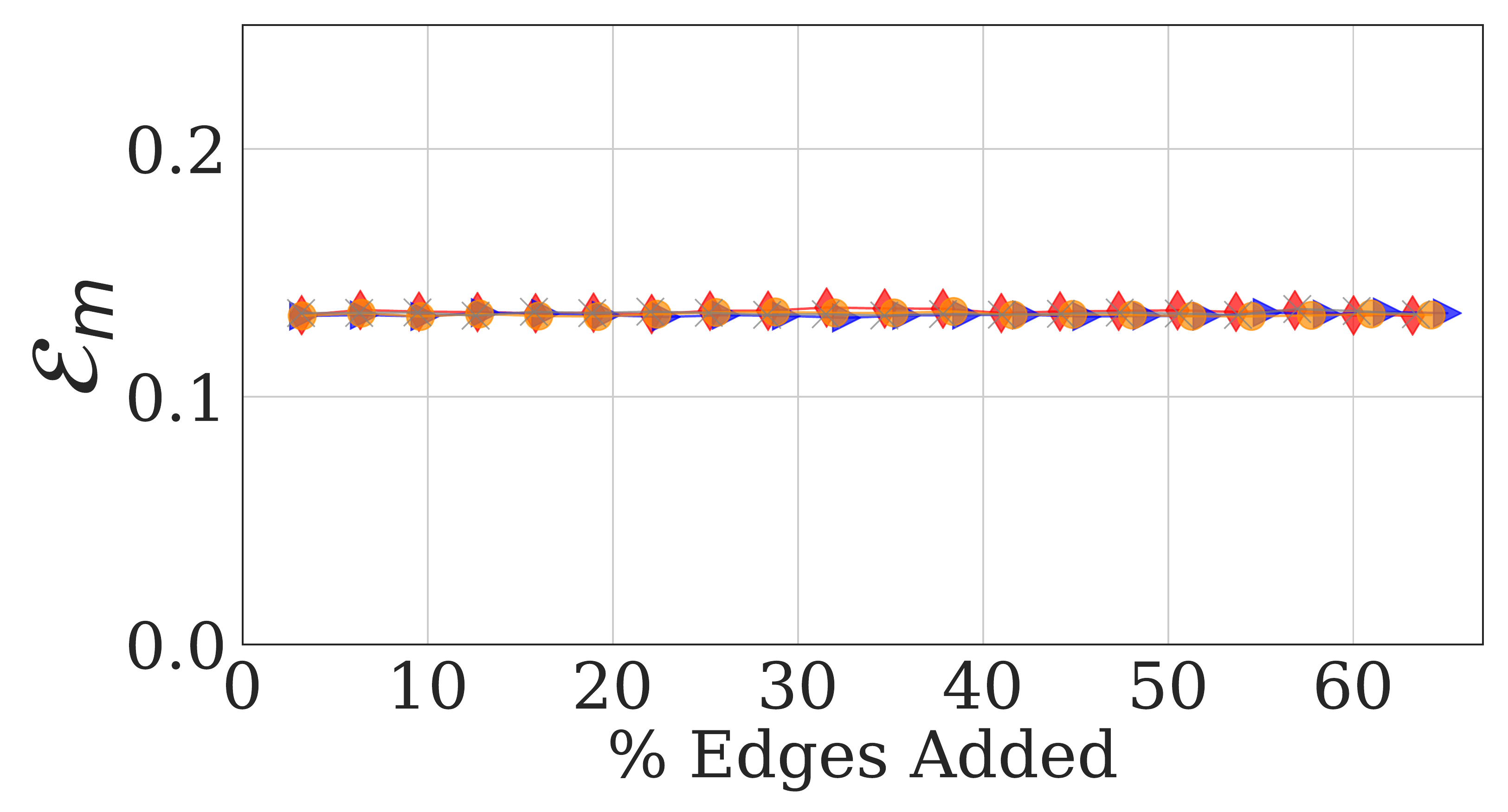} &
        \hspace{-3mm}\includegraphics[width=0.5\linewidth]{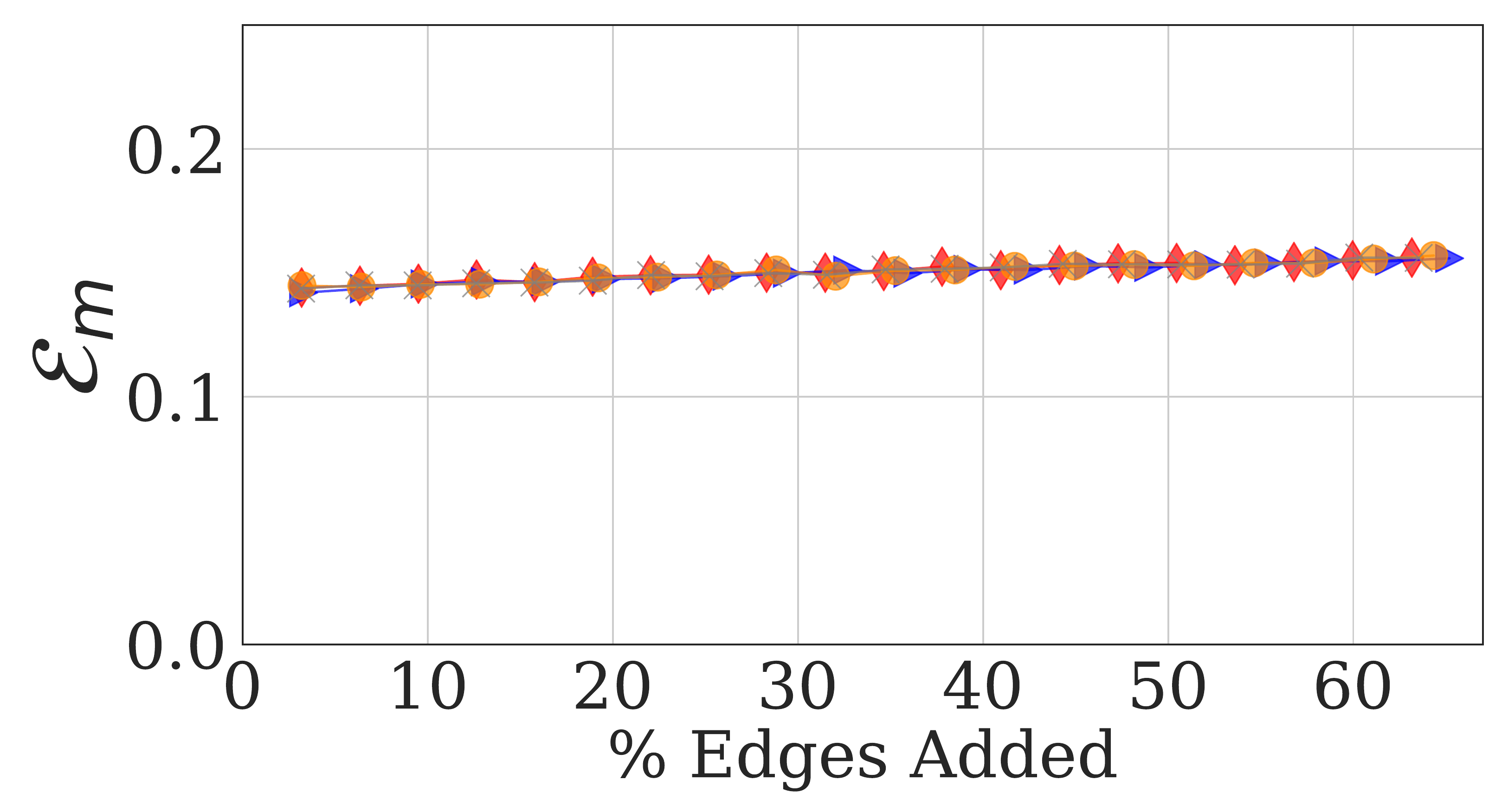}\\
        (iii) ADA & (iv) RND\\
    \end{tabular}
    \vspace{2mm}
    \caption{Exposure of minority when using different acceptance policies, running on G0.}
\label{fig:vis_recsys_fixed}
\end{figure}

In all the experiments presented so far we were adopting the B-PSB (position bias) user behaviour model.
We next analize the effect of different user behaviour models.
Figure \ref{fig:vis_recsys_fixed} reports the exposure of the minority tracked under three different policies on G0.
Each plot represents a recommender and each line in the plots represents the trend of $\mathcal{E}_{m}$ for one user behavior. For all the plots there is not such a significant difference in trends between models. This means that in circumstances where user behavior is either homogenous (B-LZY, B-PSB and B-RND)  or heterogeneous (B-MIX), and the organic growth of the network is not considered, the effect of the recommenders are consistent.

\smallskip

\mybox{mygray}{\begin{observation} The different user behaviour models do not impact the exposure in our simulations as much as the type of recommender system and the initial configuration of the network do.
\end{observation}
}

\smallskip

\subsection{Rich-get-richer effect}
After having analyzed the inequality in exposure at the group level, we now focus on the in-degree distribution at the individual level, focusing on the relationship with the popularity of the nodes (number of followers or in-degree).

As already observed in the literature, new links injected in the network can alter the inequality in the distribution of in-degree \cite{goel1}. For this reason, we study here the evolution of the rich-get-richer effect within the two groups, focusing the attention on how the in-degree of nodes is altered by the recommended output in the long-term.
We compute the Gini coefficient to analyze the level of concentration of in-degree within the two group of nodes, after each iteration. Although it was introduced in economics to measure the income or wealth inequality, Gini coefficient is widely used to measure inequalities in general \cite{ineq_univer}. It is defined as follows:
$$
G = \frac{1}{N} \left( N + 1 - 2  \frac{\sum_{i = 1}^{N}(N + 1 - i)y_i}{\sum_{i = 1}^{N} y_i}  \right).
$$

\begin{figure}[h]

    \centering
    \includegraphics[width=.9\linewidth]{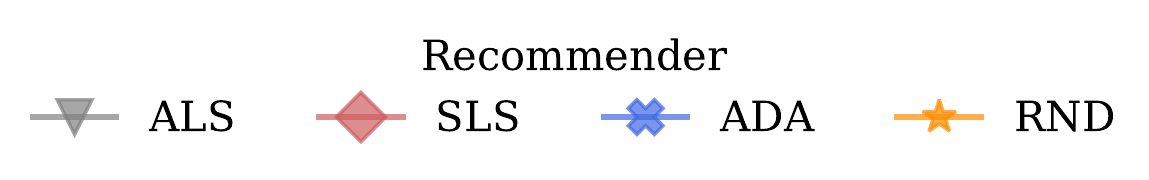}

    \begin{tabular}{cc}
    (a) Minority & (b) Majority \\

    \hspace{-3mm}\includegraphics[width=0.5\linewidth]{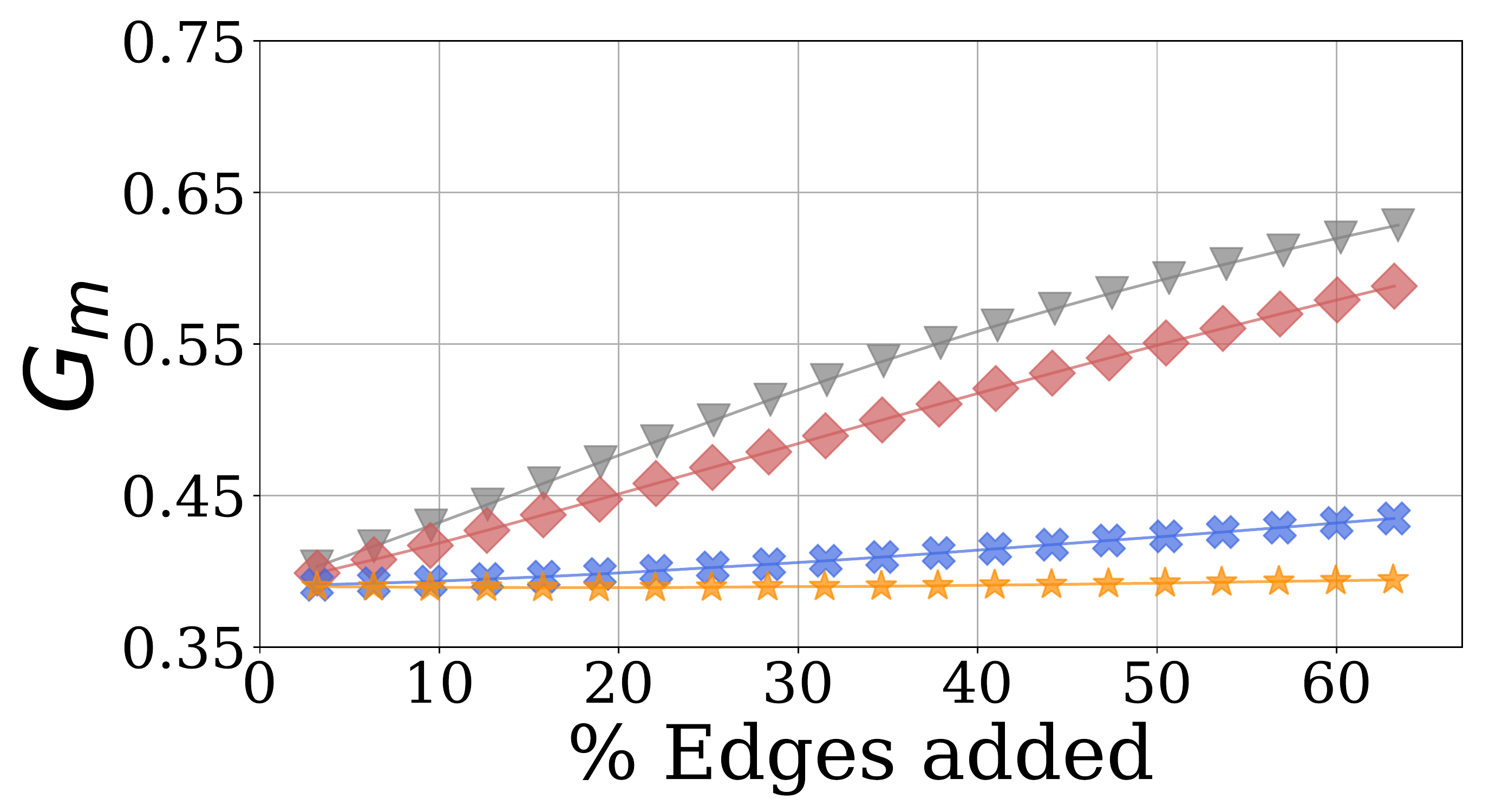} &
    \hspace{-3mm}\includegraphics[width=0.5\linewidth]{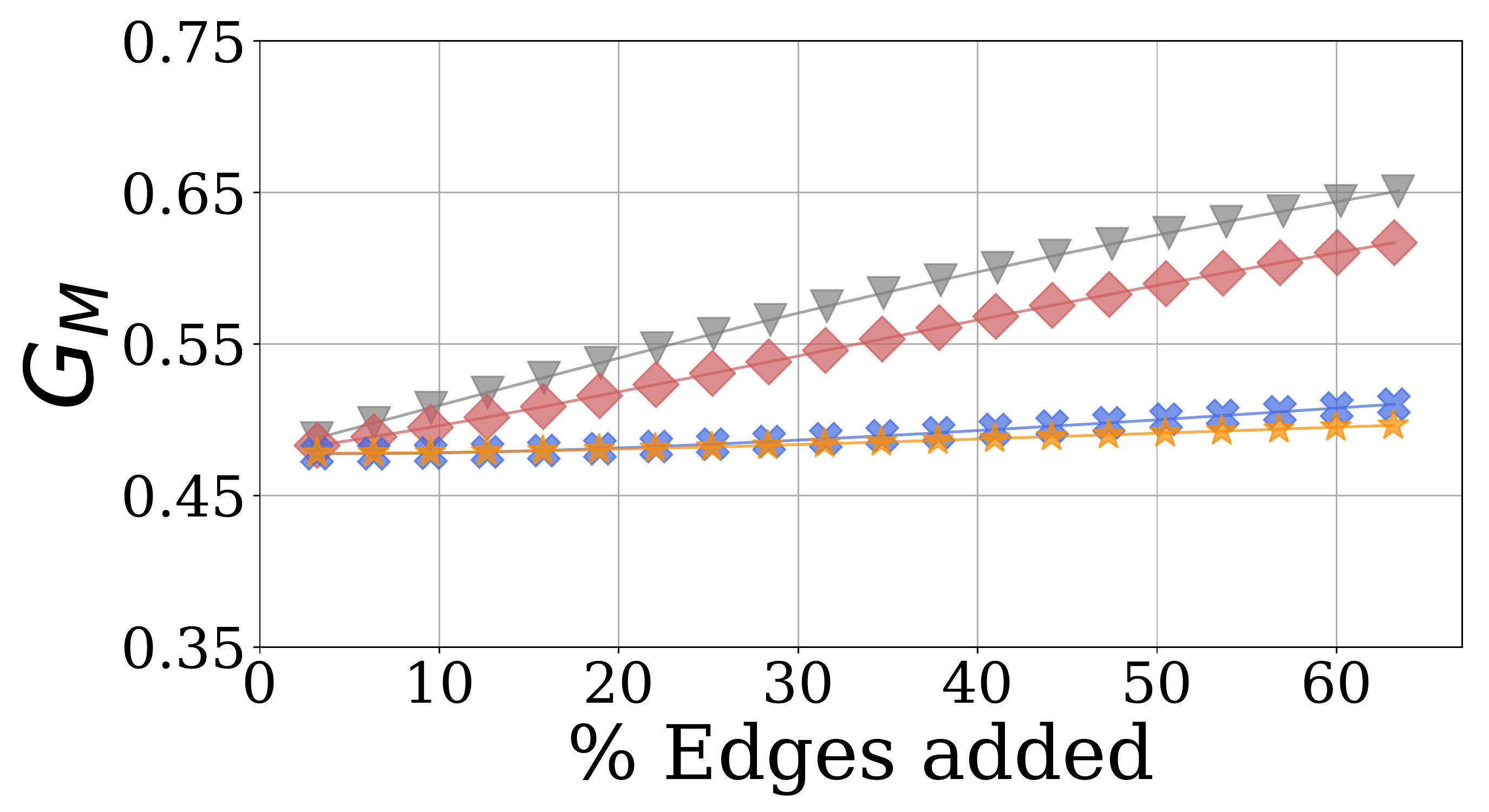}
    \end{tabular}
    \smallskip
    (i) G1
    \begin{tabular}{cc}
    \hspace{-3mm}\includegraphics[width=0.5\linewidth]{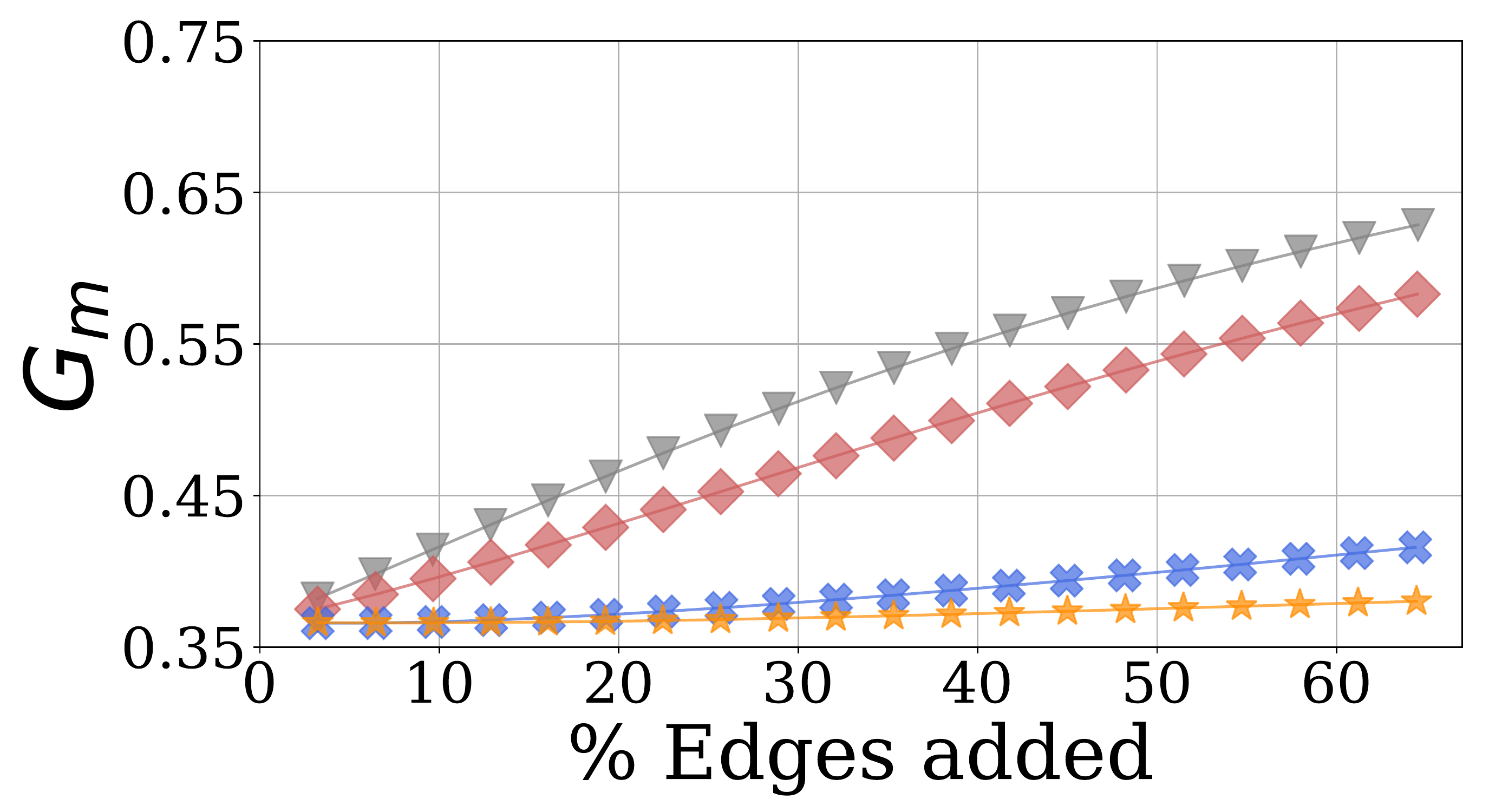} &
    \hspace{-3mm}\includegraphics[width=0.5\linewidth]{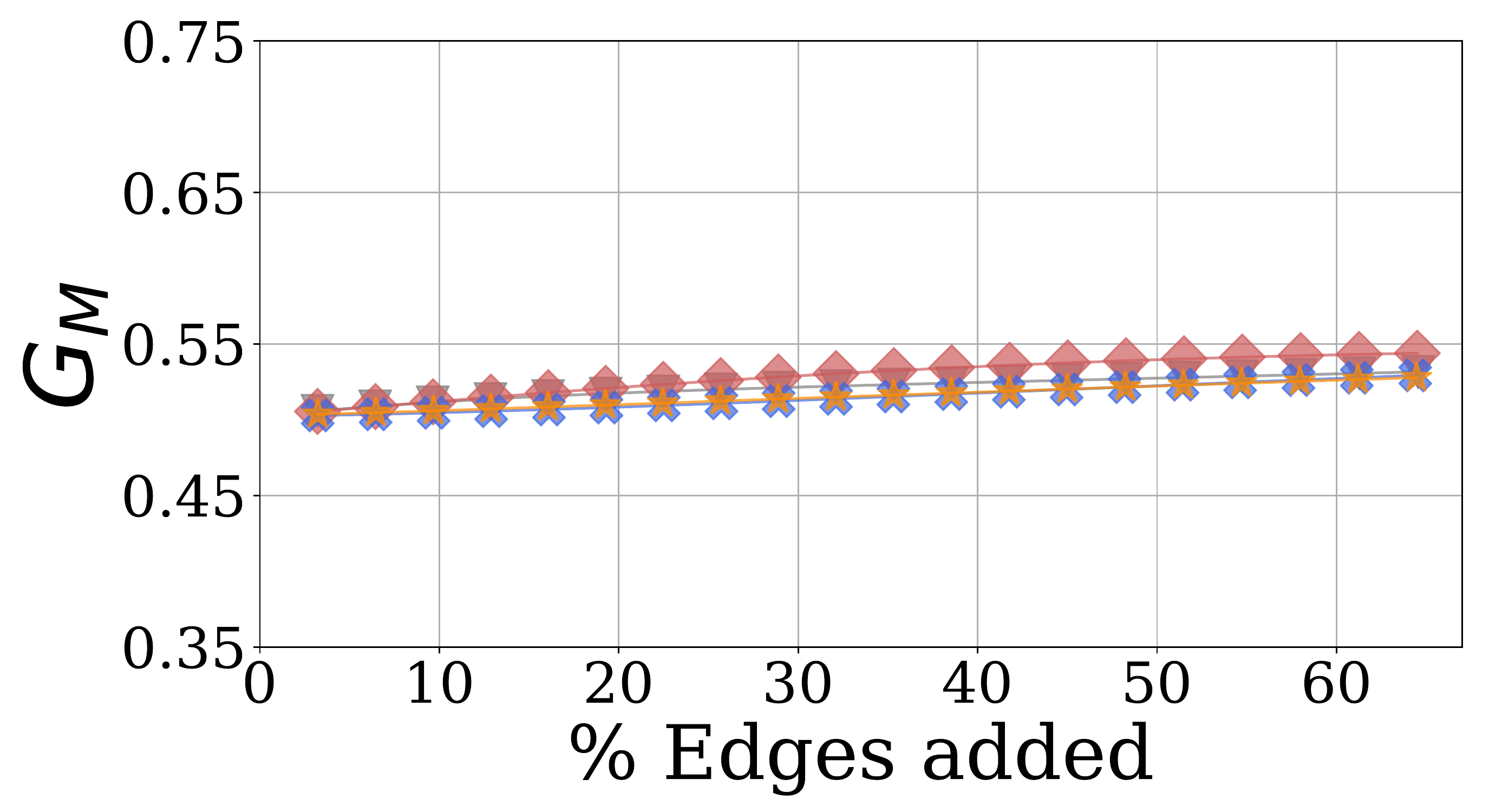}
    \end{tabular}
    \smallskip
    (ii) G2
    \begin{tabular}{cc}
    \hspace{-3mm}\includegraphics[width=0.5\linewidth]{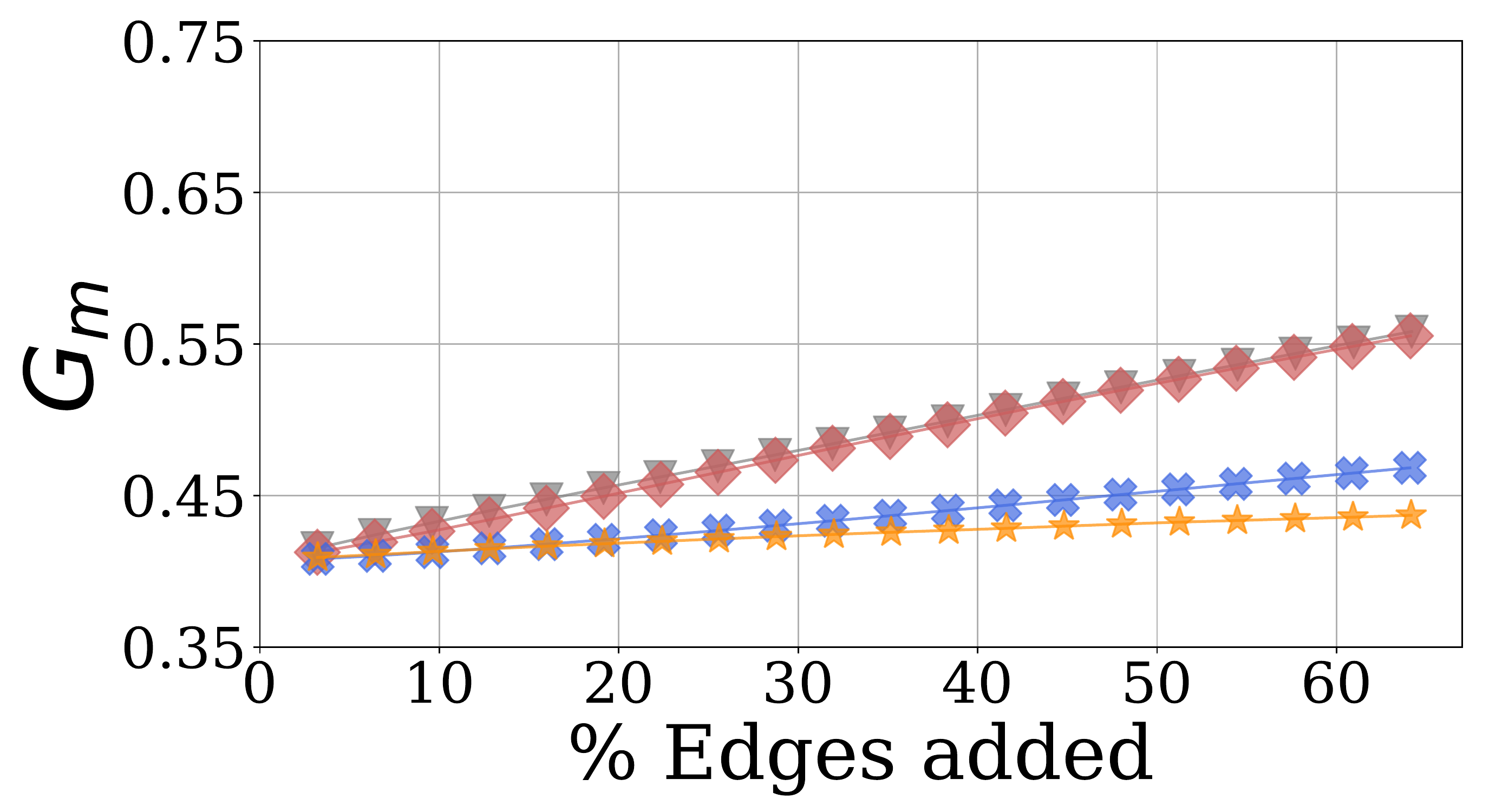} &
    \hspace{-3mm}\includegraphics[width=0.5\linewidth]{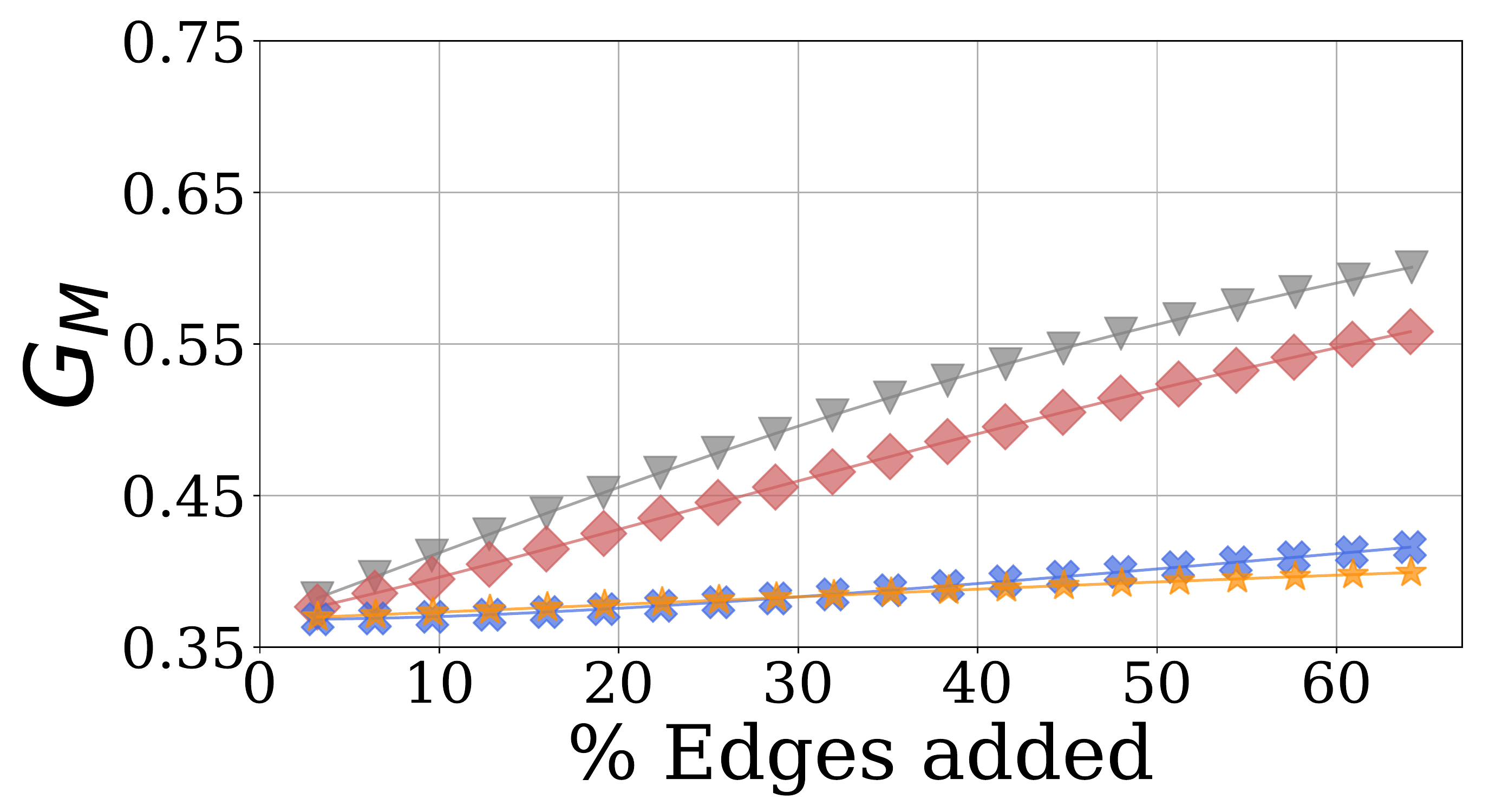}
    \end{tabular}
    \smallskip
    (iii) G3
    \begin{tabular}{cc}
    \hspace{-3mm}\includegraphics[width=0.5\linewidth]{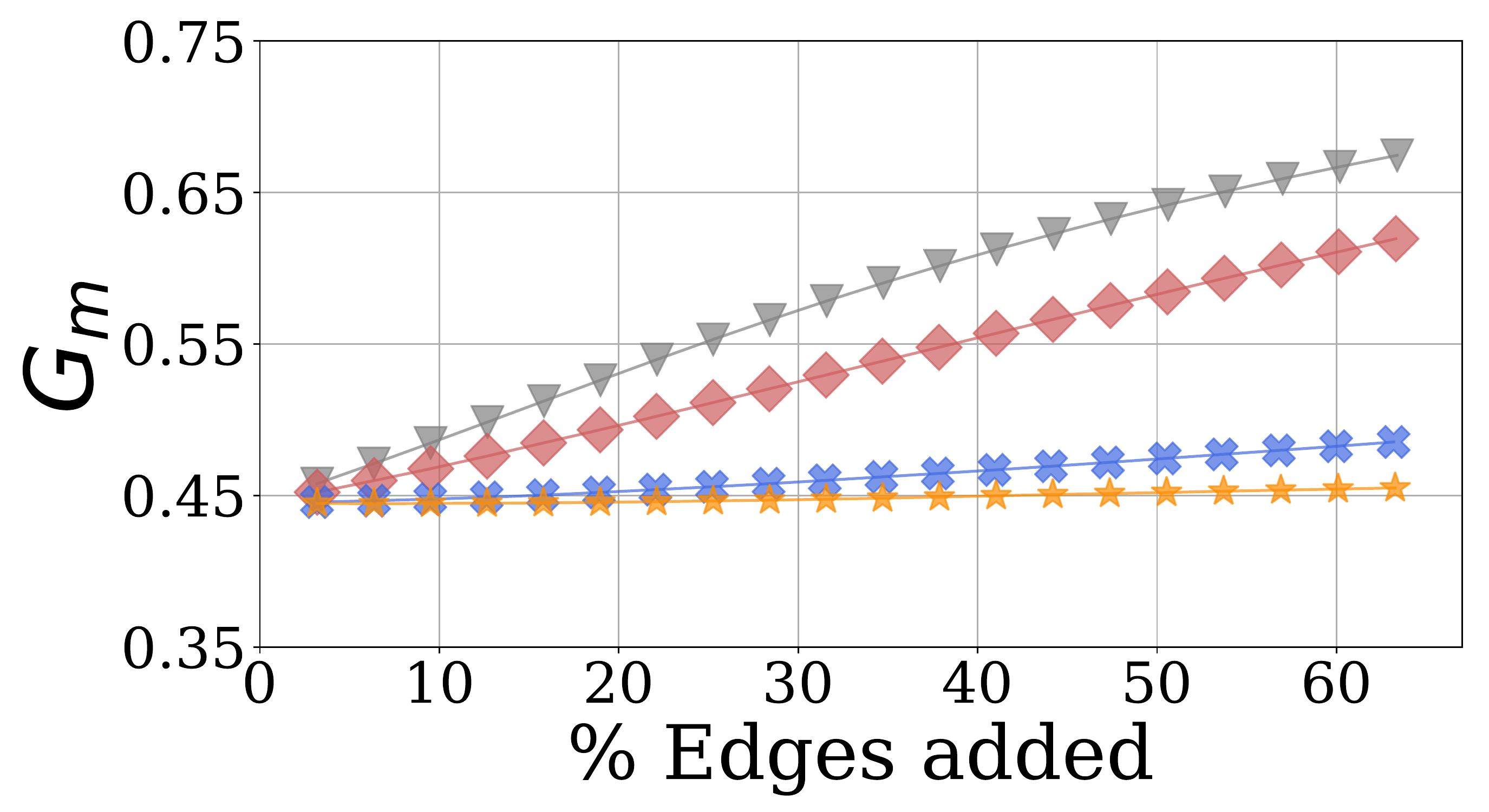} &
    \hspace{-3mm}\includegraphics[width=0.5\linewidth]{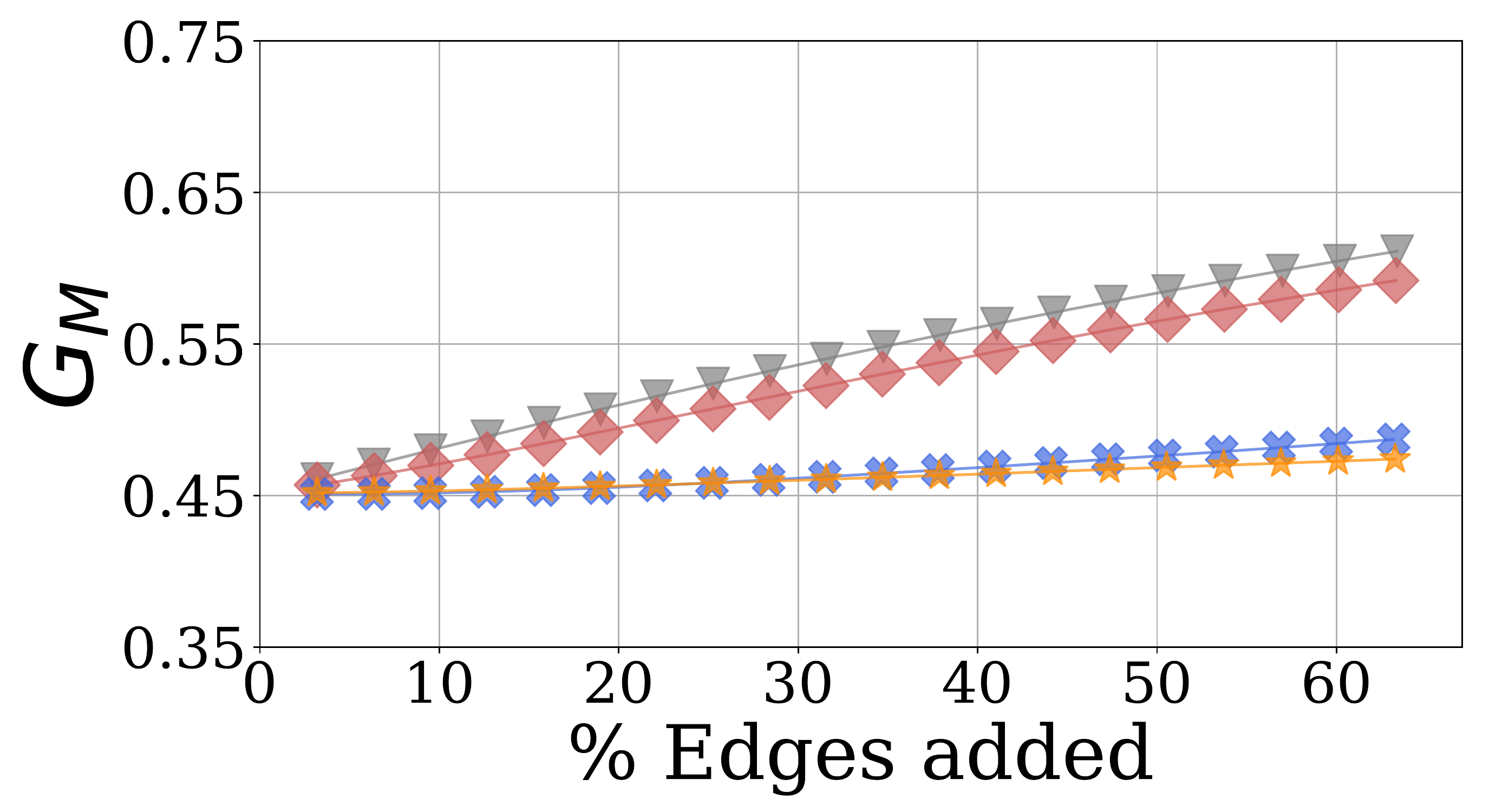}
    \end{tabular}
    (iv) G4

\caption{Gini coefficient computed on the in-degree of both minority and majority, for all the recommenders and networks, with B-PSB.}
    \label{fig:gini}
    \vspace{-3mm}
\end{figure}

In our context $N$ is the number of nodes and $y_i$ is the in-degree of the \emph{i-th} node, which has been indexed in ascending order by $y_i \leq y_{i+1}$. The index ranges from zero to one, where if all nodes receive the same amount of quantity (in-degree) then it is 0, and 1 if only one node receive the total amount. Thus, the higher the coefficient is, the higher the inequality distribution is as well.
Figure \ref{fig:gini} reports the Gini index in the long-run. Each row indicate the network, each column of plots refers to a group (minority or majority) and each line shows the Gini index after each iteration, when using all the four recommenders and one user behavior $(B-PSB)$.

For all the networks we can observe a rich-get-richer effect in both the minority and the majority class:
inequality of in-degree, as expressed by the Gini index, keeps growing, meaning that the high-degree nodes keep receiving more and more recommendations. Among the recommenders, ALS and SLS present a faster growth of the in-degree inequality, while ADA and RND are by far slower.
 When the two groups have comparable size and only one of them is homophilic (G2), the non-homophilic group gets a less severe effect.
It is also evident that when the minority is heterophilic, the more impacted group is the majority, which even if not presenting biased preferences (either homophilic or heterophilic), experiences a stronger positive trend for the growth of Gini index.

The homophily level of one group impacts, not only the inequality in in-degree distribution inside the group, but indirectly it affects also the inequality in the rest of the graph.

\begin{figure}[t]
\centering

\hspace{-5mm}\includegraphics[width=0.9\linewidth]{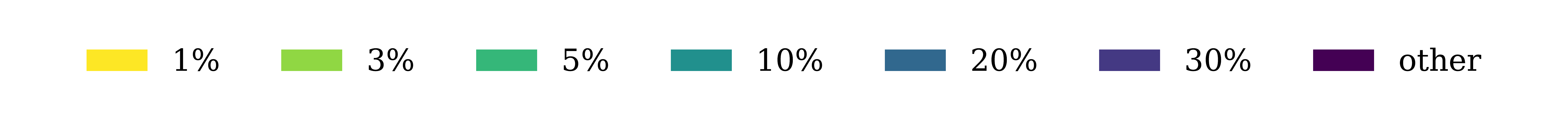}

\rotatebox[origin=c]{90}{ADA\qquad\qquad ALS\hspace{+1mm}}\begin{tabular}{cc}

(i) Minority & (ii) Majority\\
\hspace{-1mm}\includegraphics[width=0.5\linewidth]{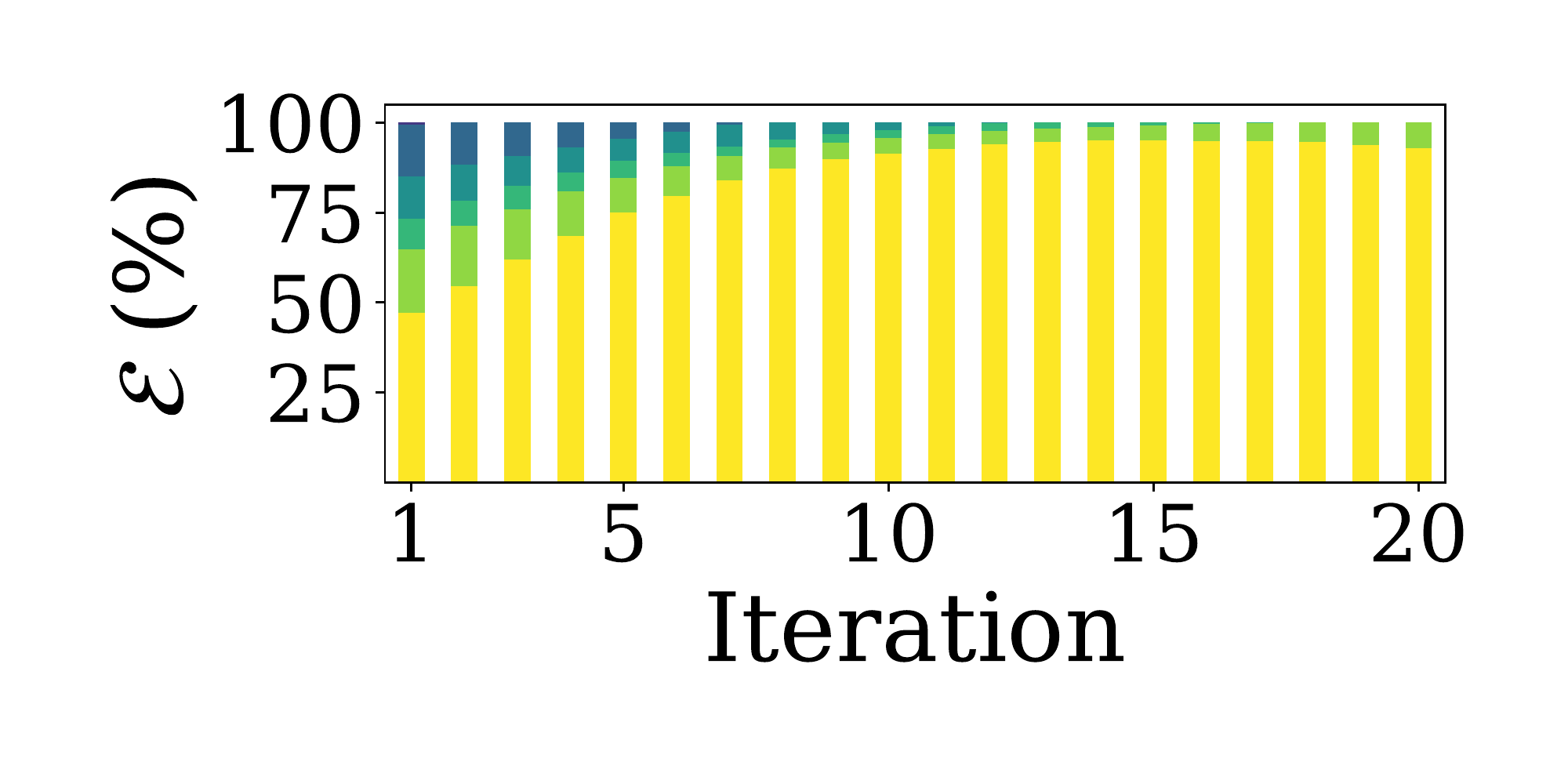}&
\hspace{-8mm}\includegraphics[width=0.5\linewidth]{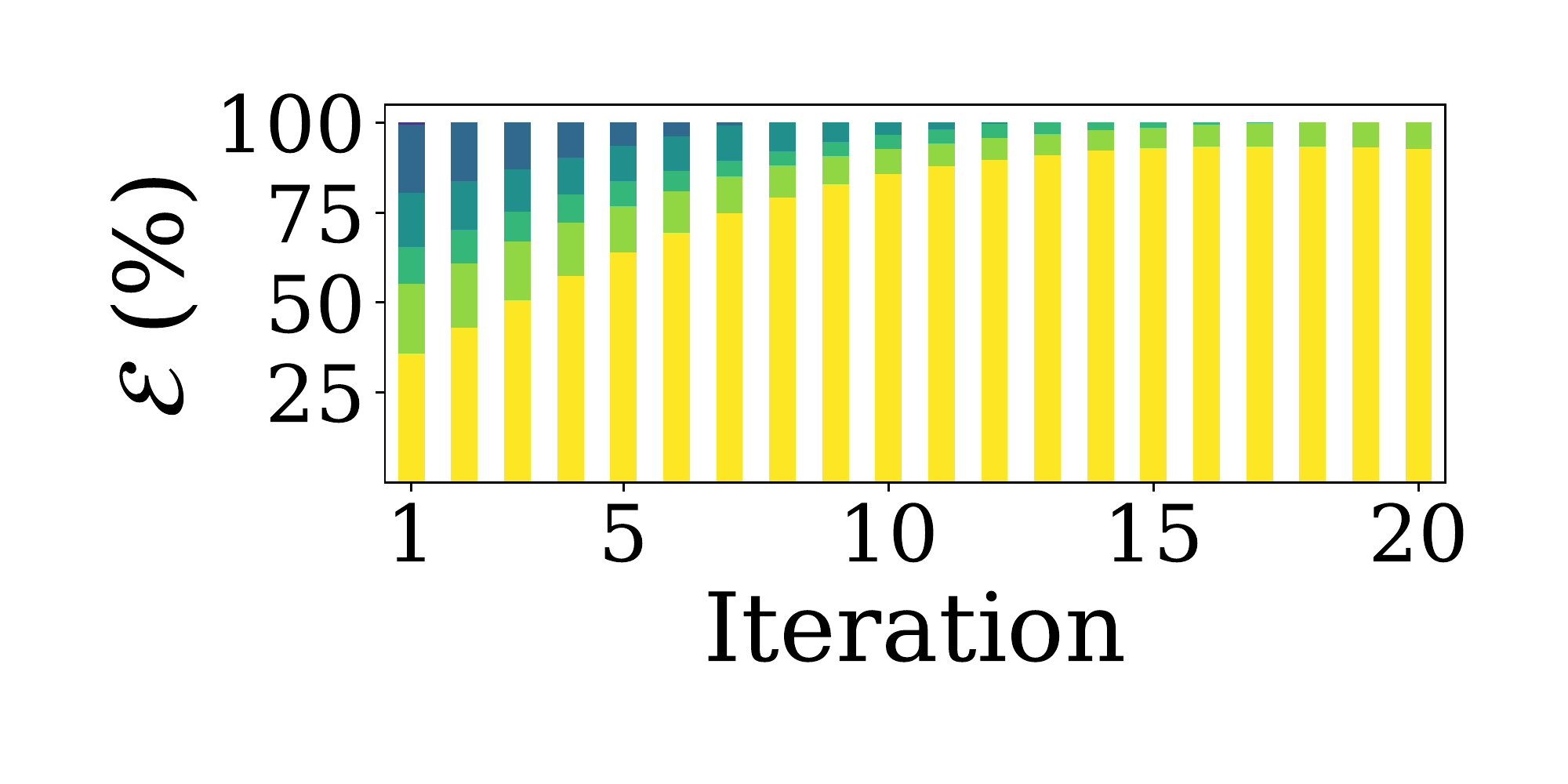}\\

\hspace{-1mm}\includegraphics[width=0.5\linewidth]{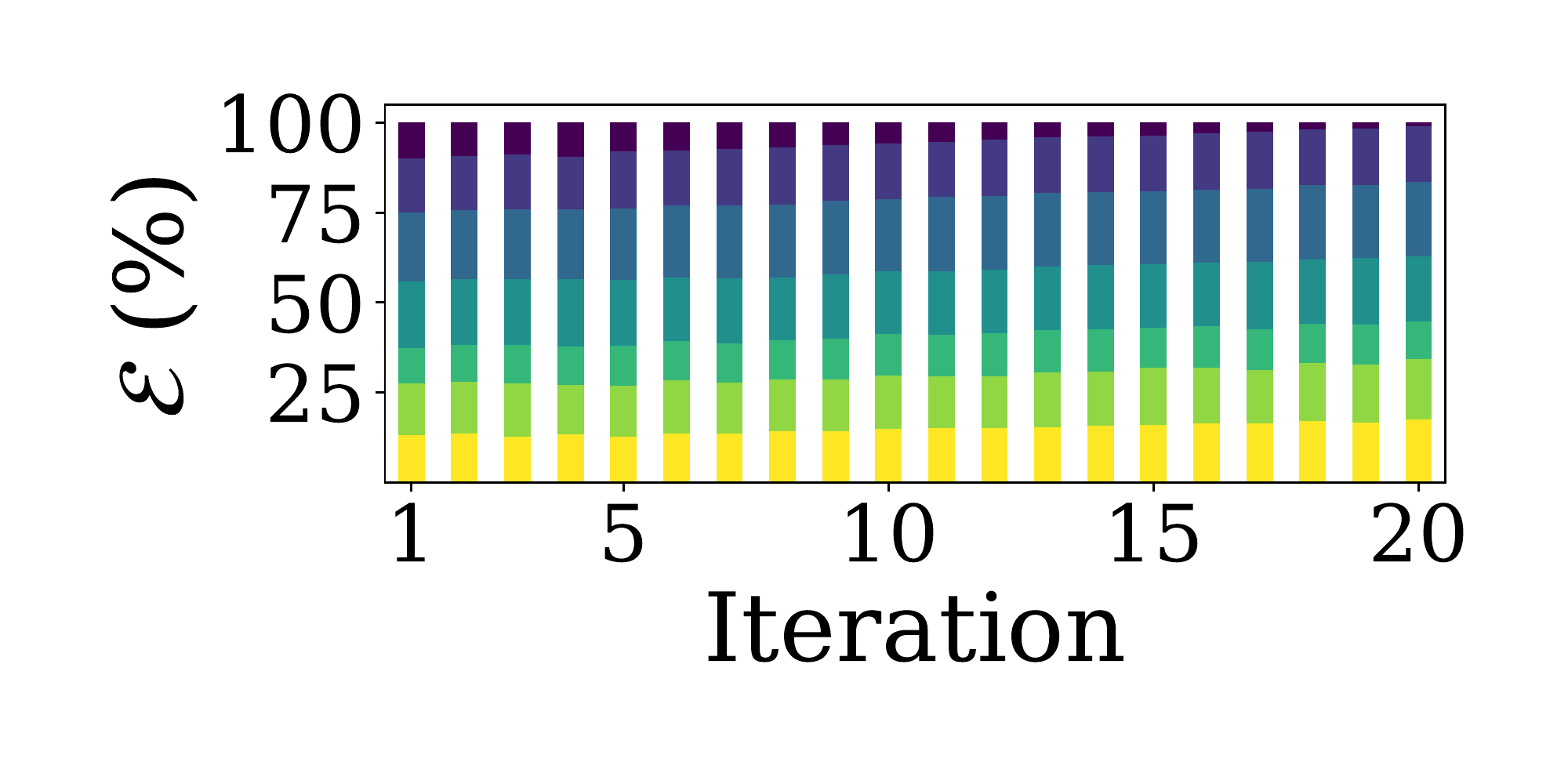}&
\hspace{-8mm}\includegraphics[width=0.5\linewidth]{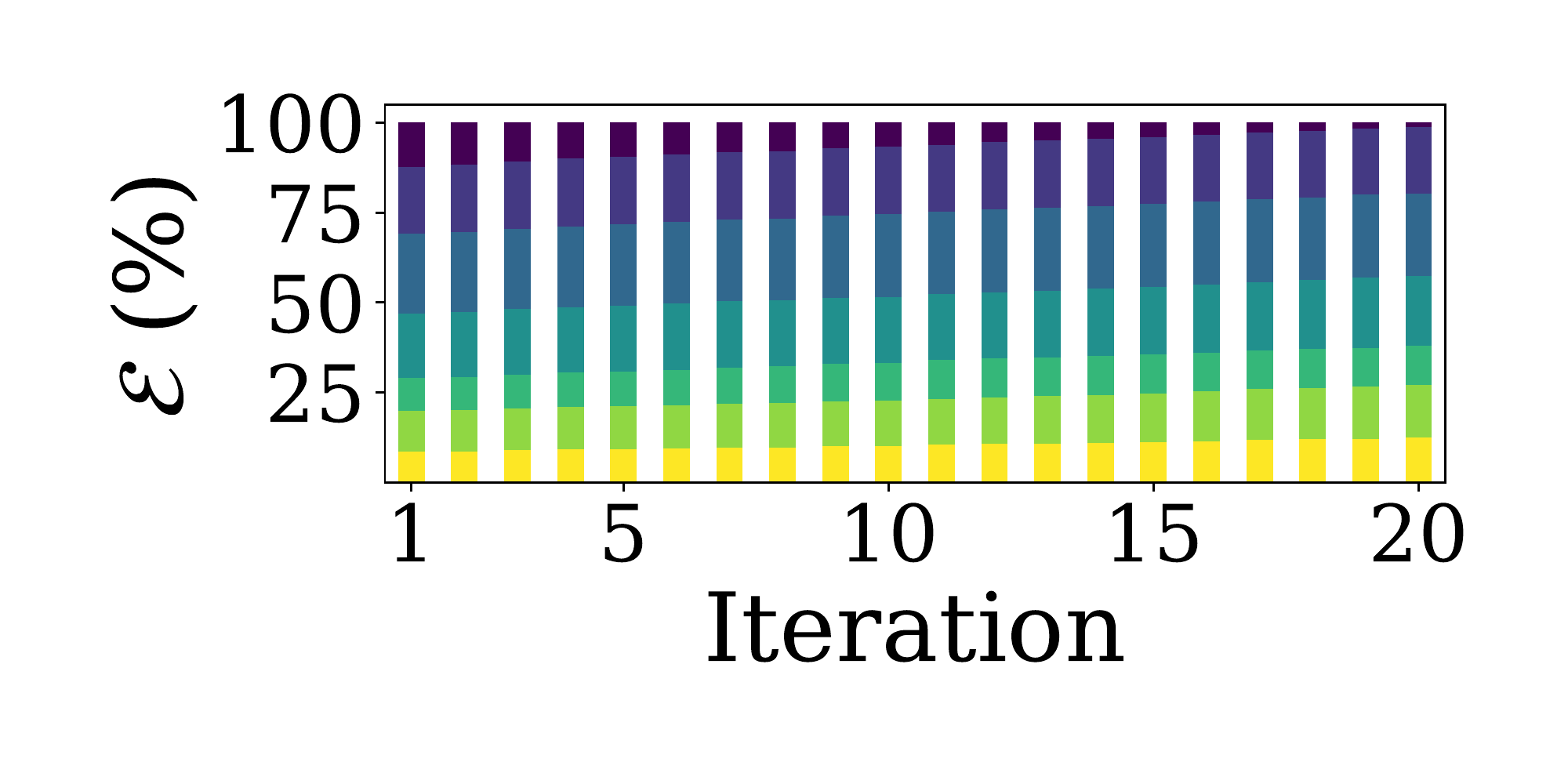}\\

\end{tabular}

(a) G1

\rotatebox[origin=c]{90}{ADA\qquad\qquad ALS\hspace{-3mm}}\begin{tabular}{cc}
\hspace{-1mm}\includegraphics[width=0.5\linewidth]{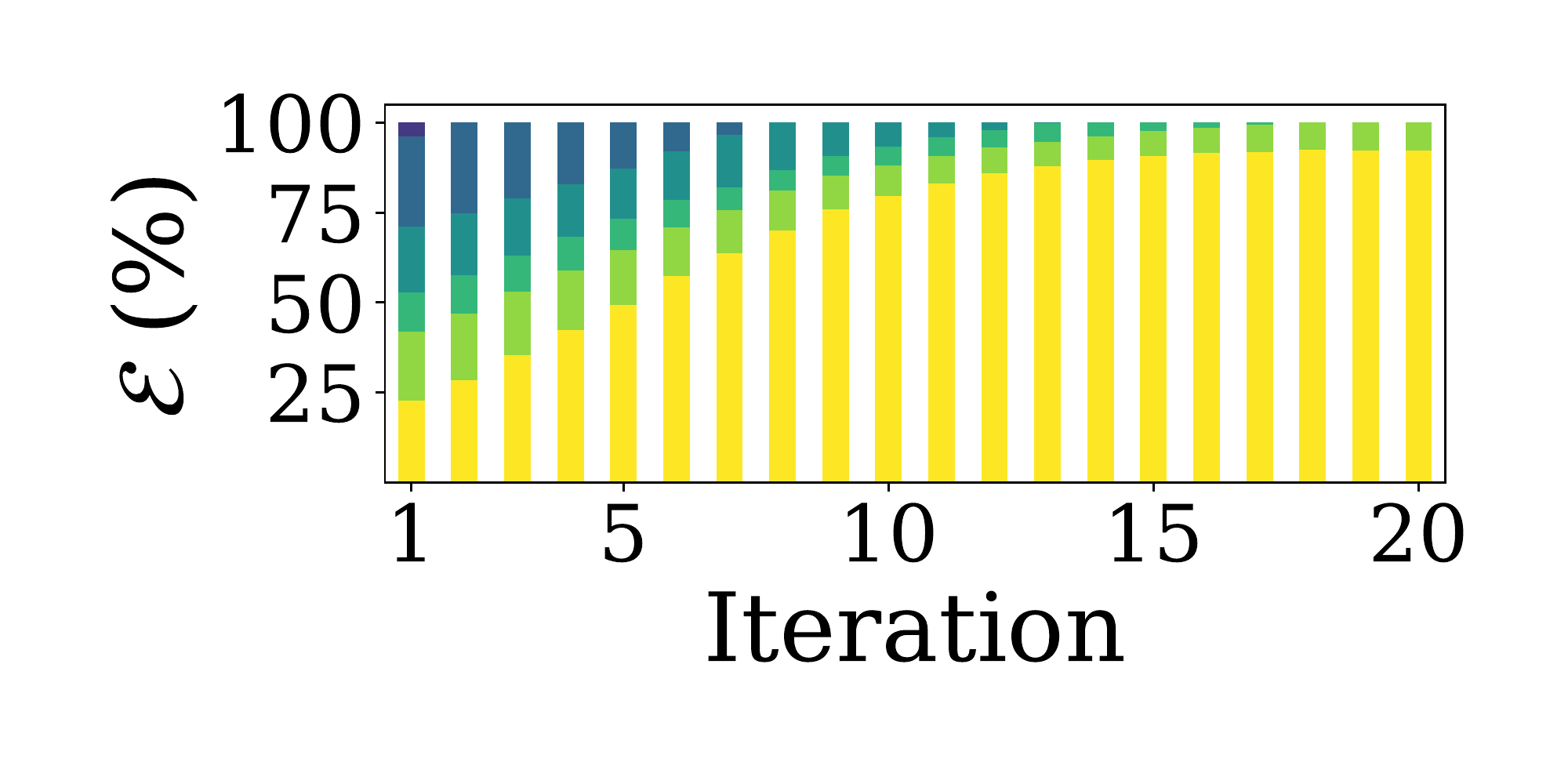}&
\hspace{-8mm}\includegraphics[width=0.5\linewidth]{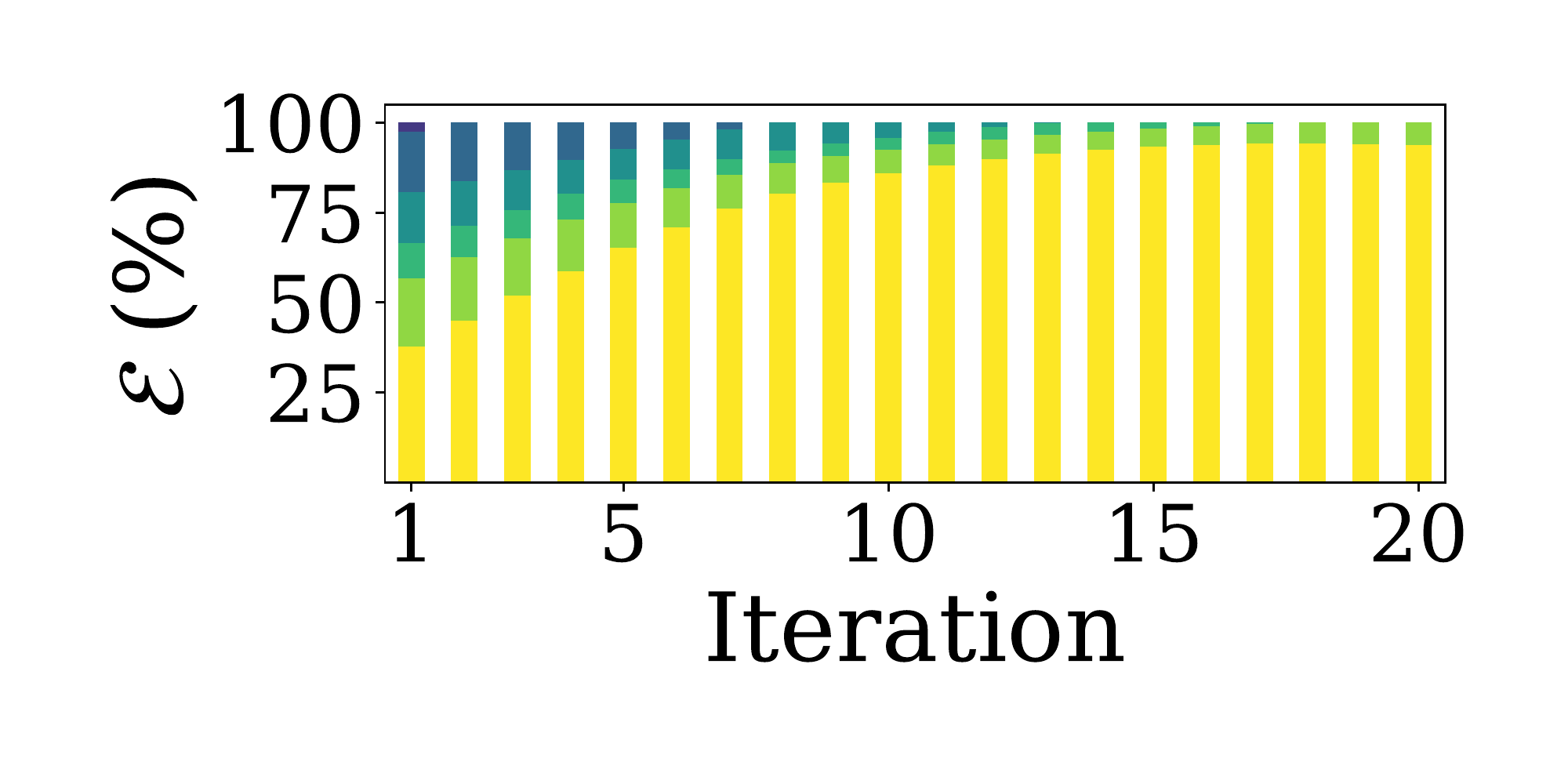}\\

\hspace{-1mm}\includegraphics[width=0.5\linewidth]{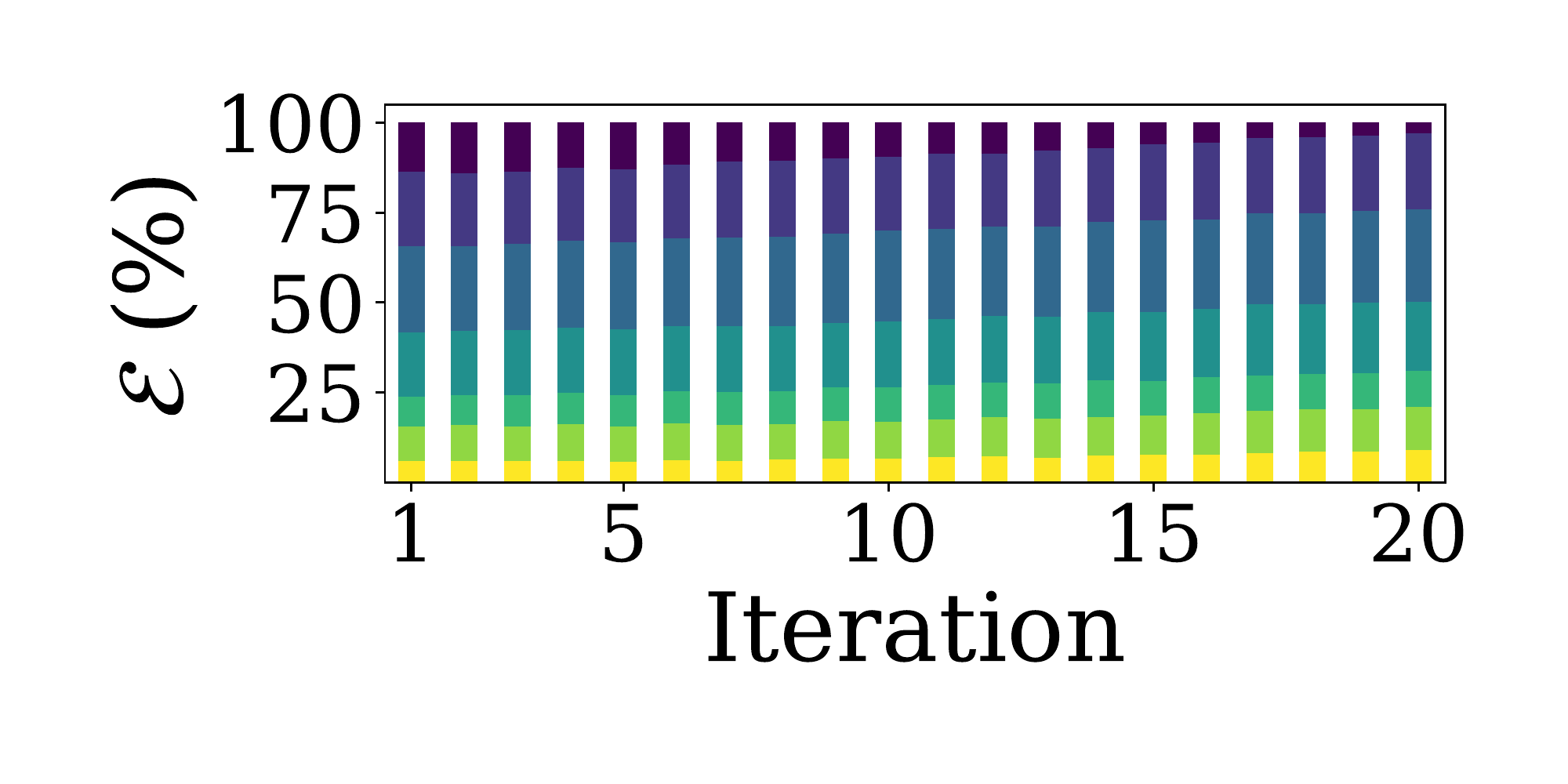}&
\hspace{-8mm}\includegraphics[width=0.5\linewidth]{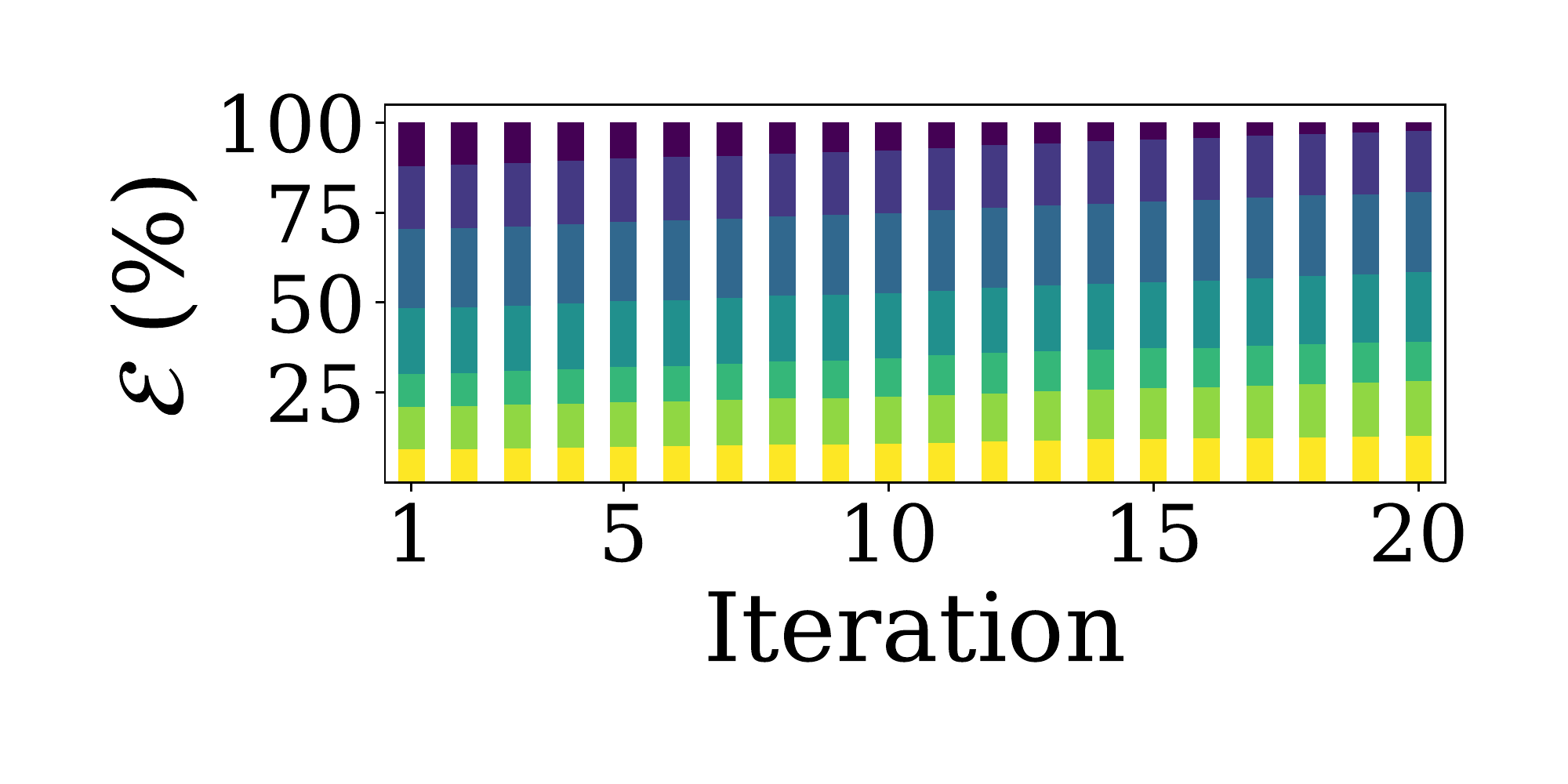}\\

\end{tabular}
(b) G3

\caption{Distribution of Exposure among nodes, where each color delimitates the \% of nodes with highest degree (best seen in color).}

\label{fig:percentile}
\vspace{-4mm}
\end{figure}

After having observed  an exacerbation of the rich-get-richer effect in the long-run, we next monitor the distribution of exposure between nodes in the two groups.
We study how subset of nodes, grouped by different in-degree, can be exposed differently in the long-run.
In Figure~\ref{fig:percentile} we show the cumulative distribution of exposure accumulated by nodes ordered by their in-degree. In particular, each bar is divided by colors, where starting from the bottom, it represents the subset of nodes having at most the correspondent in-degree. This means that, for example, the first three colors (from the bottom) represents the first $5\%$ of the nodes having the highest in-degree. On the $y$-axis we track the fraction of visibility accumulated by the nodes.

For sake of space, in Figure~\ref{fig:percentile} we focus only on two recommenders, but results are consistent also with the other two models. In particular, we have similar findings for ALS and SLS, while ADA results really close to RND.
Figure~\ref{fig:percentile} shows that, with graphs presenting either a homophilic or a heterophilic minority (G1, G3), only a subset of nodes receives most of the exposure produced. What is also evident is that after each iteration, the number of nodes getting most of the exposure become smaller, confirming some recent analytical results that point out how rankings can be biased towards few individuals getting most of the exposure \cite{germano2019few}.
Specifically, this effect, in the long-term, results faster for two specific groups: the homophilic minority in G1 and the non-homophilic majority in G3. In the first case, after only 5 iterations the nodes belonging to the top-$1\%$ acquires more than $75\%$ of exposure.
While in G3, despite the majority group not being biased, after 15 iterations, only the top-$3\%$ of nodes is recommended.
Moreover, in both graphs, ADA does not present the same increase in disparity in the long-term, but still, only a small fraction of users $(20\%)$ receives consistently the $75\%$ of the exposure in both groups.

\smallskip

\mybox{mygray}{\begin{observation} When the minority presents non-neutral preferences (either homophilic or heterophilic), ALS and SLS can increase disparity in both exposure and in-degree: a small subset nodes benefits in terms of exposure by the injection of new links, and those are also the ones with highest degree. The cardinality of this subset of nodes becomes smaller after each iteration.
\end{observation}
}

\smallskip

\subsection{Model evaluation}

\begin{figure}[t!]
    \centering

    \begin{tabular}{c}
        \hspace{-5mm}\includegraphics[width=.8\linewidth]{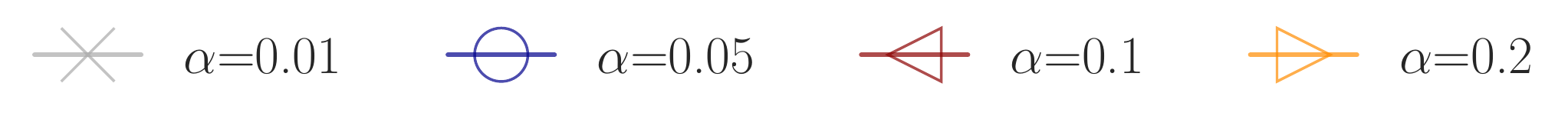} \\

        \hspace{-5mm}\includegraphics[width=0.8\linewidth]{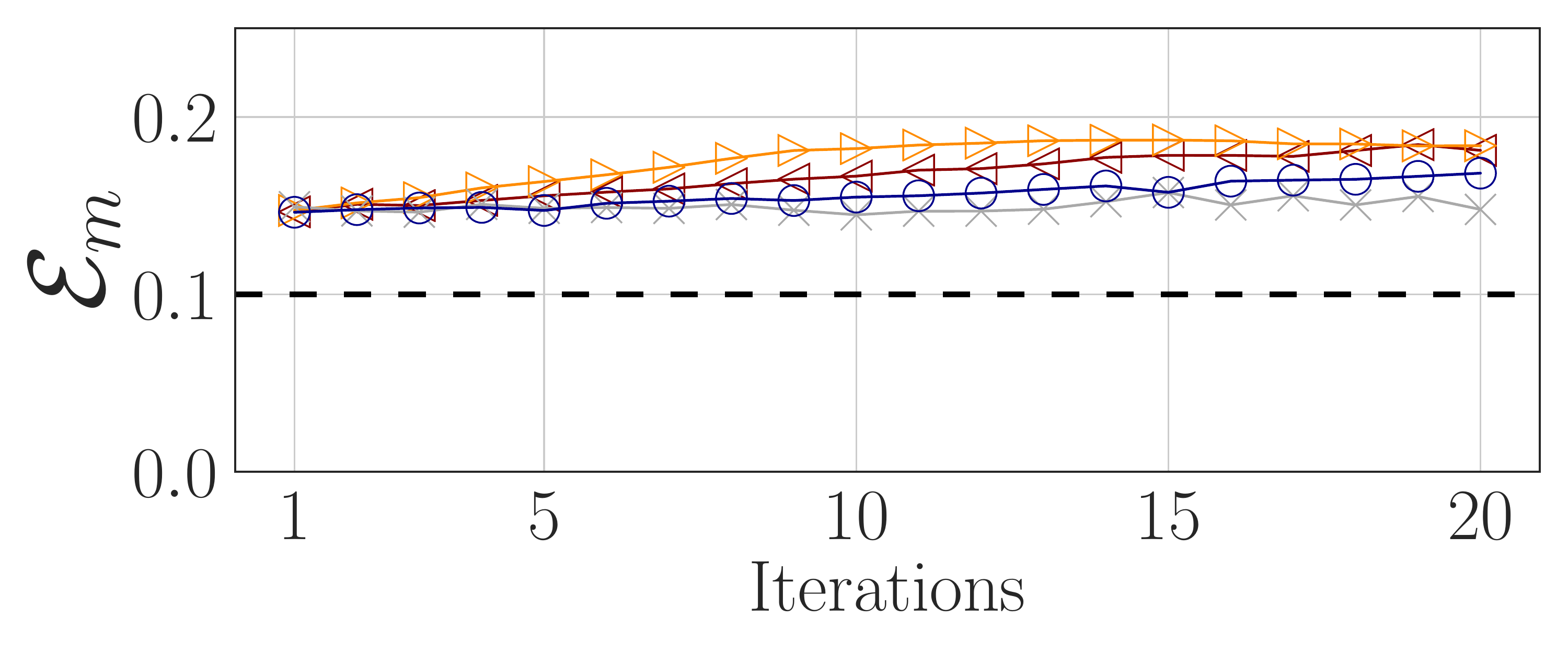} \\
       (a) Level of Sparsity\\

        \hspace{-5mm}\includegraphics[width=0.8\linewidth]{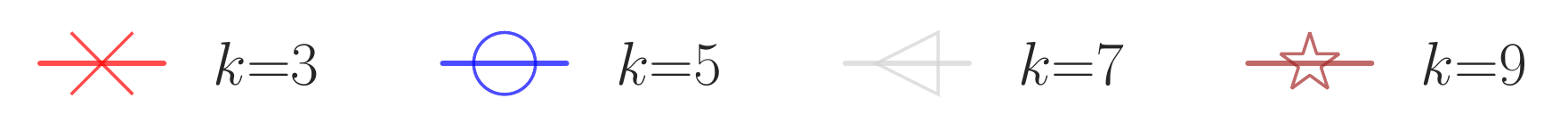}\\
        \hspace{-5mm}\includegraphics[width=0.8\linewidth]{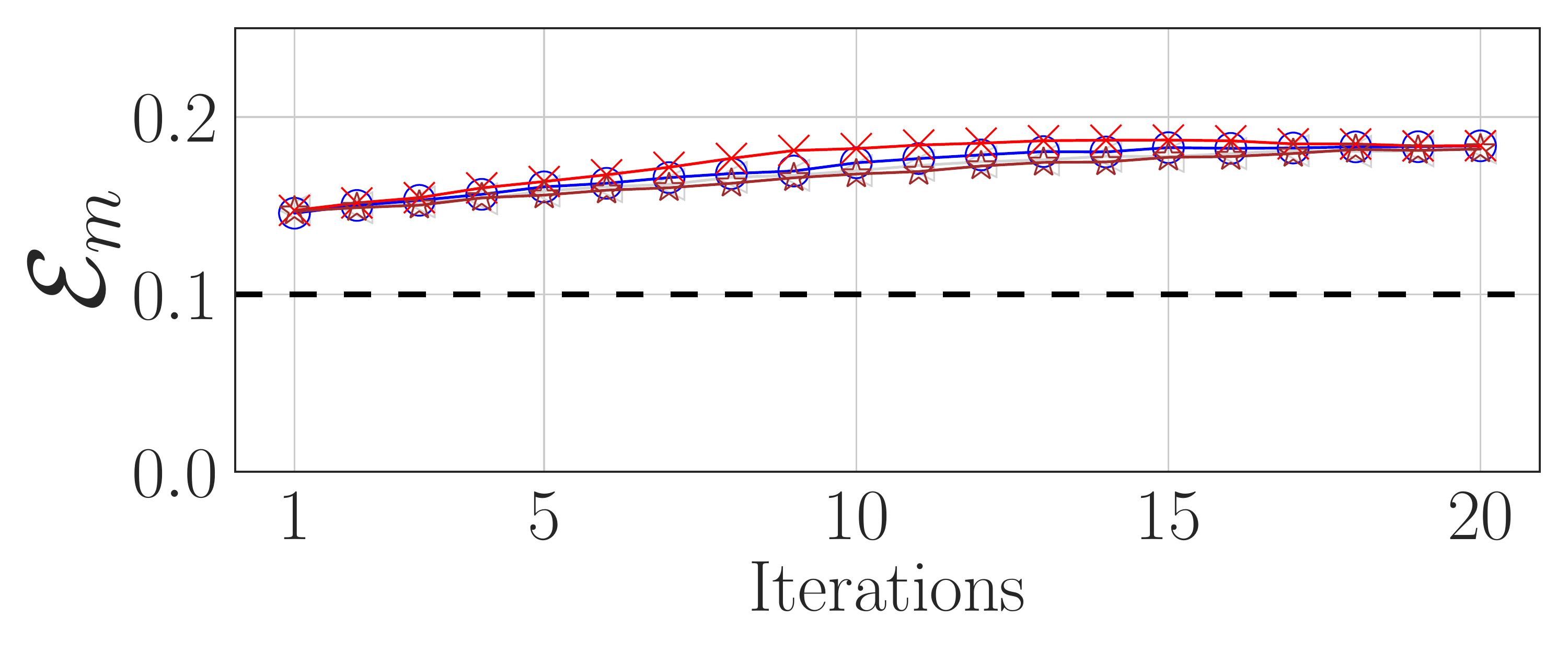}\\
        (b) Length of Top-\emph{k}

    \end{tabular}

\caption{Testing different values of $\alpha$ and $k$ on G1, with B-PSB.}
    \label{fig:diagnostics}

\end{figure}

In the experiments seen so far, we used the same configuration of $k$ and $\alpha$. To analyze how those input parameters may affect the simulation outcome, we next produce configurations presenting more sparse interactions (smaller $\alpha$) and longer lists of recommendations (larger $k$). In Fig. \ref{fig:diagnostics} we present the exposure of the minority for G1, simulating the user behavior using B-PSB. In the first figure (Fig. \ref{fig:diagnostics}a) we tune different values of $\alpha$, which the smaller, the less the number of users sampled to submit new recommendations. The plot shows how the exposure tends to growth, as expected, but with a slower pace. This parameter can be tuned looking at interactions generated by the social media platform over time. Here the length of recommendations is fixed to $k=3$.
In the case of longer lists of recommendations, the length of recommendation output impacts even less on the final output (Fig. \ref{fig:diagnostics}b). The effect observed with the smaller recommendation list ($k=3$) presents a trend close to all the other configurations. In this case, $\alpha$ is fixed to the usual value of $0.2$.

\medskip

\section{Limitations and Implications}

The main goal of this work is to improve our comprehension of the long-term consequences on the disparate exposure of a minority in the recommendations provided by people recommender systems in a social network. In this endeavor we need to take care of the interplay between the recommender algorithm, user behavior in accepting the recommendations, and pre-existing conditions in the network (e.g., the existence of an homophilic minority). However, disentangling the consequences of these single components remains challenging. Hence, we next discuss some limitations of our study, proposing potential extensions for future investigation.

Our analysis shows how the initial level of homophily within a subpopulation in the graph can drive exposure inequalities that grow over time: this is obtained without considering organic growth of the network (i.e., new links are created only if recommended by the algorithm) and assuming a homogenous user behavior for accepting or rejecting the link recommendations. In the future, we plan to extend our framework, including organic network growth as part of the simulation and allowing the users to have heterogenous behaviors. In this way, we may study how homophily evolving over time may impact the recommendation output. Moreover, the user behavior models analyzed here do not consider homophily as a potential factor. In our future effort we plan to investigate how homophily can impact user choices when accepting or rejecting algorithmic recommendations.


Our work analyzes the impact of human biases, such as homophilic behavior, and link recommender algorithms on the disparate exposure of a minority at the level of the whole network. A more fine-grain analysis at the mesoscale level of communities or subgraphs might be useful to better understand the phenomena at play.

Our work raises critical observations about the long-term consequences of algorithms in online social networking platforms and hints the need to design algorithms which keeps in consideration existing biases and aim at mitigating the long-term consequences, instead of exacerbating them.
The challenge is the usual trade-off between generating recommendations which are still relevant for the users, while being able to mitigate the disparate exposure between groups even after repeated round of recommendations.

Moreover, designers of social network platforms, may take benefit of using simulation framework to infer potential harmful scenarios: the framework we present in this work can be easily adapted to test new recommendation algorithms, e.g., to shed the light on the consequences of introducing new features, before the deployment. 

Finally, the results of the present study highlight the urgency to include link recommendation algorithms among the key elements when modeling network dynamics.


\medskip 

\section*{Acknowledgments}
Francesco Fabbri was partially supported by the Catalan Government through the funding grant ACCI\'{O}-Eurecat (Project PRIVany-nom). Francesco Bonchi acknowledges support from Intesa Sanpaolo Innovation Center. 
Carlos Castillo was partially supported by the HUMAINT programme (Human Behaviour and Machine Intelligence), Joint Research Centre, European Commission, and from "la Caixa" Foundation (ID 100010434), under the agreement LCF/PR/PR16/51110009.
The funders had no role in study design, data collection
and analysis, decision to publish, or preparation of the manuscript.

\end{document}